\documentclass[12pt,draftcls, peerreview, onecolumn]{IEEEtran}

\usepackage[usenames]{color}
\usepackage{url}
\usepackage{cite}     
\usepackage{graphicx}  
\usepackage{epsf}
\usepackage{epsfig}
\usepackage{subfigure}
\usepackage{amsmath} 
\usepackage{algorithmic}
\usepackage{array}
\usepackage{amsfonts}
\usepackage{xcolor}

\graphicspath{%
    {converted_graphics/}% inserted in the PCTeX.
    {C:/Users/IBK/Desktop/Task2_converted_graphics/}% inserted in the PCTeX.
}
\begin{document}

% correct bad hyphenation here
\hyphenation{op-tical net-works semi-conduc-tor}


\title{Performance of a Threshold-based WDM and ACM for FSO Communication between Mobile Platforms in Maritime Environments}

\author{Jae-Eun Han,~\IEEEmembership{Member,~IEEE,}
        Sung Sik Nam,~\IEEEmembership{Senior Member,~IEEE,}
        Duck Dong Hwang,~\IEEEmembership{Member,~IEEE,}
        and~Mohamed-Slim Alouini,~\IEEEmembership{Fellow Member,~IEEE}%
%\thanks{This work was supported in part by the National Research Foundation of Korea (NRF) grant (NRF-2021R1F1A1047271)}%
\thanks{J.-E. Han and S. S. Nam (corresponding author) are with the department of Electronic Engineering, Gachon University, Korea (E-mail: $\langle$ssnam$\rangle$@gachon.ac.kr). D. Hwang (co-corresponding author) is with the Department of Electronics and Communication Engineering, Sejong University, South Korea.
M.-S. Alouini is with the Computer, Electrical and Mathematical Science and Engineering Division (CEMSE), King Abdullah University of Science and Technology (KAUST), Thuwal, Makkah Province, Saudi Arabia (E-mail: slim.alouini@kaust.edu.sa).}}

\markboth{S.S.Nam \MakeLowercase{\textit{et al.}}}%
{Shell \MakeLowercase{S.S.Nam\textit{et al.}}}



%\tableofcontents
%\newpage

\maketitle


\begin{abstract}
In this study, we statistically analyze the performance of a threshold-based multiple optical signal selection scheme (TMOS) for wavelength division multiplexing (WDM) and adaptive coded modulation (ACM) using free space optical (FSO) communication between mobile platforms in maritime environments with fog and 3D pointing errors.
Specifically, we derive a new closed-form expression for a composite probability density function (PDF) that is more appropriate for applying various algorithms to FSO systems under the combined effects of fog and pointing errors.
We then analyze the outage probability, average spectral efficiency (ASE), and bit error rate (BER) performance of the conventional detection techniques (i.e., heterodyne and intensity modulation/direct detection). 
The derived analytical results were cross-verified using Monte Carlo simulations. 
The results show that we can obtain a higher ASE performance by applying TMOS-based WDM and ACM and that the probability of the beam being detected in the photodetector increased at a low signal-to-noise ratio, contrary to conventional performance.
Furthermore, it has been confirmed that applying WDM and ACM is suitable, particularly in maritime environments where channel conditions frequently change.
\end{abstract}


% \fi
%\IEEEpeerreviewmaketitle

\section{Introduction}
In the case of communication between mobile platforms (e.g., ship-to-ship) in maritime environments, interest in adopting free-space optical (FSO) communication in maritime environments is increasing owing to an increase in the demand for rapid data exchange among mobile platforms \cite{Xiao}. However, maritime environments pose unique challenges to FSO communication owing to fog and pointing errors due to the nature of maritime environments. In maritime environments, fog, particularly from water droplets and various types of aerosols, can cause major attenuation of the laser beam and reduce visibility by several meters in a worst-case scenario (e.g., 480 dB/km) \cite{Olga}. Ultimately, this can significantly affect the performance and reliability of the FSO communication \cite{Maged2016}. Another problem in maritime environments is that because of the characteristics of maritime environments, the continuous movement of the platform makes it difficult for the laser beam to be orthogonal to the photodetector (PD) plane, resulting in pointing errors that require non-orthogonality to be considered in the FSO communication system \cite{Marzieh}. These pointing errors can cause additional attenuation and signal distortion, further degrading performance. Overall, FSO communication systems have the potential to revolutionize mobile platform communications in maritime environments; however, the combined effects of fog and pointing errors in maritime environments can significantly degrade the reliability and performance of FSO communications.
It is necessary to analyze the complex effects of fog and pointing errors statistically to solve these problems, and then, based on this, develop an algorithm that can mitigate performance degradation that may occur due to these effects and provide reliable communication in maritime environments.

In terrestrial environments, improving the FSO performance by applying wavelength division multiplexing (WDM) or adaptive coded modulation (ACM) has already been extensively studied \cite{Mazin,Naazira,Anuranjana,Abu,Abisayo,Sujon,Amit,Chun,E}, \cite{Sayed,Anshul,Y,Mehdi,K}, respectively. 
These technologies have proven to be effective in mitigating the effects of Log-normal turbulence and pointing errors in terrestrial environments \cite{Sung2017}.
Maritime FSO systems, however, must operate under uniquely unstable conditions influenced by atmospheric moisture (or water vapor) and sea level fluctuations.
To cope with the rapidly changing channel conditions in such an environment, applying WDM and ACM is more important than applying land-based FSO systems. 
When WDM is applied, multiple optical signals can be multiplexed and transmitted on a single medium, and data can be transmitted with minimum attenuation by selecting the optimal wavelength according to the channel conditions, even in unstable channel
environments with severe changes, thereby enabling stable communication. 
ACM selects an appropriate channel code and modulation scheme based on the channel state that can increase the average spectral efficiency (ASE) while satisfying the required bit error rate (BER). 
By applying this technology, optimal communication based on channel status can be achieved, even in an unstable channel environment. 
Therefore, it is necessary to respond to an unstable channel environment by applying WDM and ACM to maritime FSO systems. 
The application of this technology enables stable communication, thereby contributing to the
performance improvement in maritime FSO systems.

This study applied WDM and ACM technologies to improve the performance of FSO communication in rapidly changing maritime channel environments, focusing on overcoming the combined challenges of fog and pointing errors. 
When applying WDM to maritime FSO systems, not all the multiple beams generated by WDM can provide an acceptable signal-to-noise ratio (SNR) owing to the unstable channel environment. 
Therefore, in this study, a signal selection method that selects only effective beams that provide a received SNR above a certain pre-selected threshold, that is, a threshold-based multiple optical signal selection scheme (TMOS), is applied to reduce the implementation complexity and enable efficient communication \cite{Sung2017}.
Applying TMOS-based WDM and ACM to mitigate performance degradation owing to the combined effects of fog and pointing errors in FSO communication among maritime mobile platforms, a comprehensive statistical analysis of the system's performance, considering the intricate effects of fog and pointing errors, is central to this research.  
No study has statistically analyzed the system performance while considering these complex effects. Therefore, in this paper, we first statistically analyzed the complex effects of fog and pointing errors to derive a composite probability density function (PDF) in a closed format based
on the proposed system model with widely applied heterodyne detection (HD) and intensity modulation/direct detection (IM/DD) techniques. 
Although the channel model underwent statistical analysis, as documented in \cite{Kug2022}, this study holds significance for the field by deriving a new closed-form expression for the composite PDF, which is particularly suitable for statistical performance analysis when applying various algorithms to maritime FSO systems under the combined effects of fog and pointing errors in maritime environments. 
This analysis contributes considerably to understanding the potential of these technologies to enhance the reliability and efficiency of FSO communication between mobile platforms in maritime environments. 
Furthermore, it offers substantial insight into performance analysis when various algorithms are applied based on the newly derived composite PDF results.

\section{System and channel models}
\label{systemmodel}

\begin{figure}[ht] % float placement: (h)ere, page (t)op, page (b)ottom, other (p)age
  \centering
  % file name: C:/Users/IBK/Desktop/Task2_converted_graphics/mainfigure.eps
  \includegraphics[width=13.0 cm]{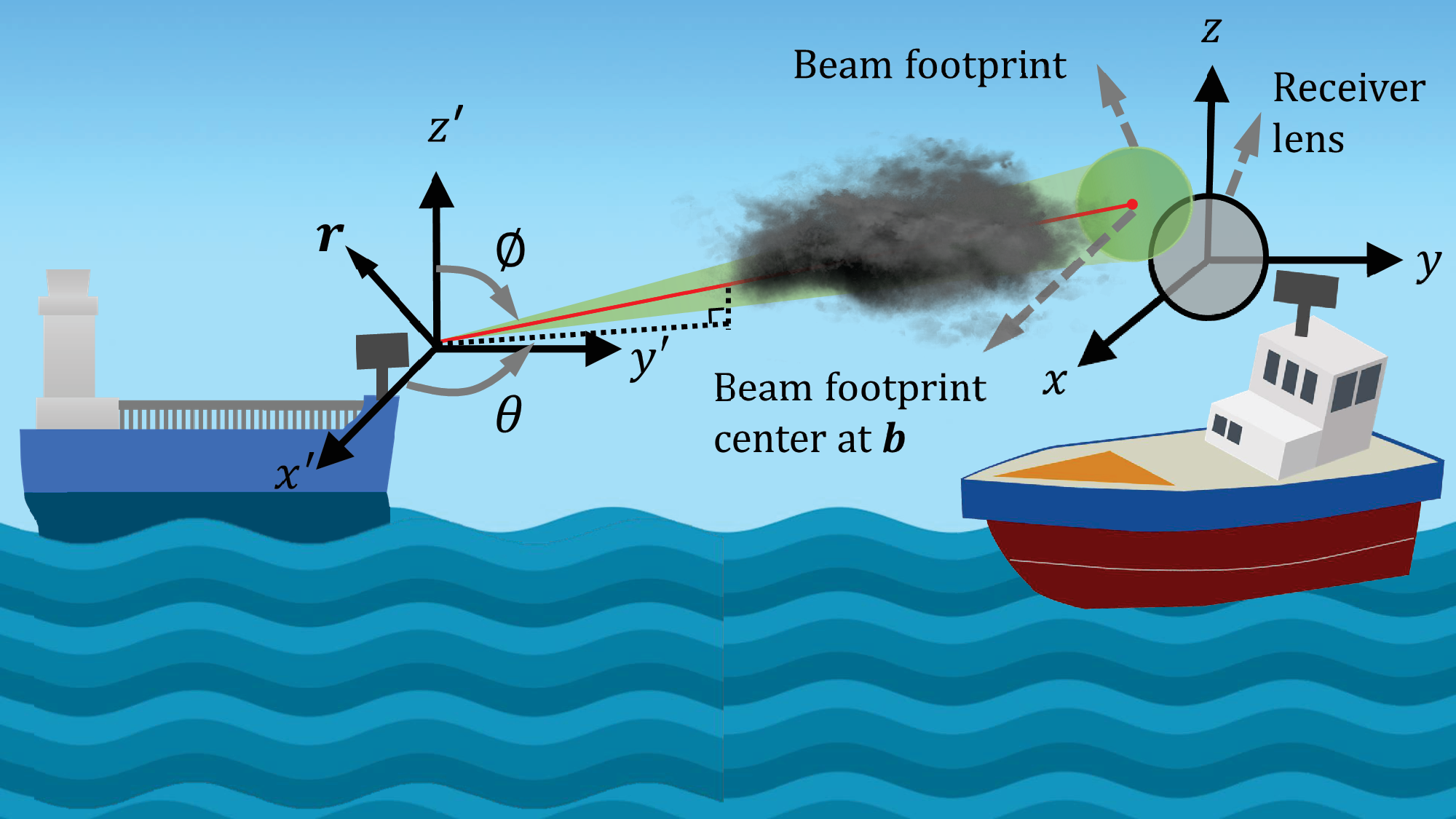}
  \caption{System and channel models of FSO communication between mobile platforms in maritime environments.}
  \label{fig:main-figure}
\end{figure}

We consider FSO communication between mobile platforms in maritime environments, as shown in Fig. \ref{fig:main-figure}.
Here, because fog and pointing errors are major factors in performance degradation, the irradiance can be modeled as 
 $I = {I_a}{I_p}$ where ${I_a}$ and ${I_p}$ are fog and pointing errors, respectively.
To reduce the effects of these two factors and increase the FSO performance,
we considered a threshold-based rate-adaptive $N$ multidimensional trellis coded M-QAM for WDM to FSO communication as \cite{Sung2017}.
Specifically, because fog and pointing errors are both random variables, it is assumed that multiple wavelengths experience the same channel condition when applying a block-fading channel that models slowly varying fading.
For slowly varying fading (symbol period $\ll$ channel's coherence), during the guard time, at the receiver, the link quality of each beam was estimated and compared with the predefined threshold $\gamma_{T}$ and then fed back to the transmitter.
Based on this information, during transmission, the transmitter transmits only beams above $\gamma_{T}$.
With these selected beams, we considered a rate-adaptive $N$ multidimensional trellis, coded as M-QAM.
Using the largest possible modulation order while maintaining a predefined target BER, the number of transmitted bits per symbol interval can be maximized such that the obtainable ASE approximates the maximum ASE.
With these assumptions, we statistically analyzed the characteristics of FSO communication between mobile platforms in maritime environments using both HD and IM/DD techniques based on our system and channel models.
In the following section, the statistical model of the foggy channel and pointing error are described.

\subsection{Foggy channel}
\begin{table} 
\centering
\caption{Values of parameters for different types of fog}
\label{fog}
\begin{tabular}{ccccc}
\noalign{\smallskip}\noalign{\smallskip}\hline
Fog type & Dense & Thick & Moderate & Light \\
\hline
$V(m)$ & 0-50 & 50-200 & 200-500 & 500-1000 \\
$k$ & 36.05 & 6.00 & 5.49 & 2.32 \\
$\beta$  & 11.91 & 23.00 & 12.06 & 13.12 \\
\hline
\label{fog}
\end{tabular}
\end{table}
In terrestrial environments, conventional mobile platform-based FSO communication considers the vertical channel between the fixed-mobile platform or the horizontal channel between the mobile platform; therefore, it is difficult to directly apply the statistical model of fog, as given in \cite{Maged2016ICT}.
However, because FSO communication between mobile platforms in maritime environments consider the form of a horizontal channel, the conventional statistical model of fog suggests that \cite{Maged2016ICT} can be used directly.
Based on \cite{Maged2016ICT}, using the Beer-Lambert law, which explains the relationship between the propagation path and signal attenuation, the foggy channel state is expressed as follows:
\begin{equation}\label{ia}
{I_a} = \exp \left( { - \alpha l/4.343} \right),
\end{equation}
where $l$ is the propagation link length in km and $\alpha$ is the signal attenuation random variable in dB/km.
The probability distribution function (PDF) of $I_a$ defined in \cite{Maged2016ICT} can be written as follows:
\begin{equation} \label{iapdf}
\begin{array}{l}
{f_{{I_a}}}({I_a}) = \frac{{{z^k}}}{{\Gamma (k)}}{\left[ {\ln \left( {\frac{1}{{{I_a}}}} \right)} \right]^{k - 1}}{I_a}^{z - 1}, \quad 0 < {I_a} \le 1,
\end{array}
\end{equation}
where $z = 4.343/\beta l$ and $\Gamma \left(  \cdot  \right)$ denotes the Gamma function \cite[Eq. (8.310.1)]{table}. Parameters $k$ and $\beta$ are the shape and scale parameters, respectively, and their related values for different fog densities are listed in Table \ref{fog}.

\begin{figure}[ht] % float placement: (h)ere, page (t)op, page (b)ottom, other (p)age
  \centering
  % file name: C:/Users/IBK/Desktop/Task2_converted_graphics/mobile-platform.eps
  \includegraphics[bb=0 0 902 635,width=14.4cm,height=10.1cm,keepaspectratio]{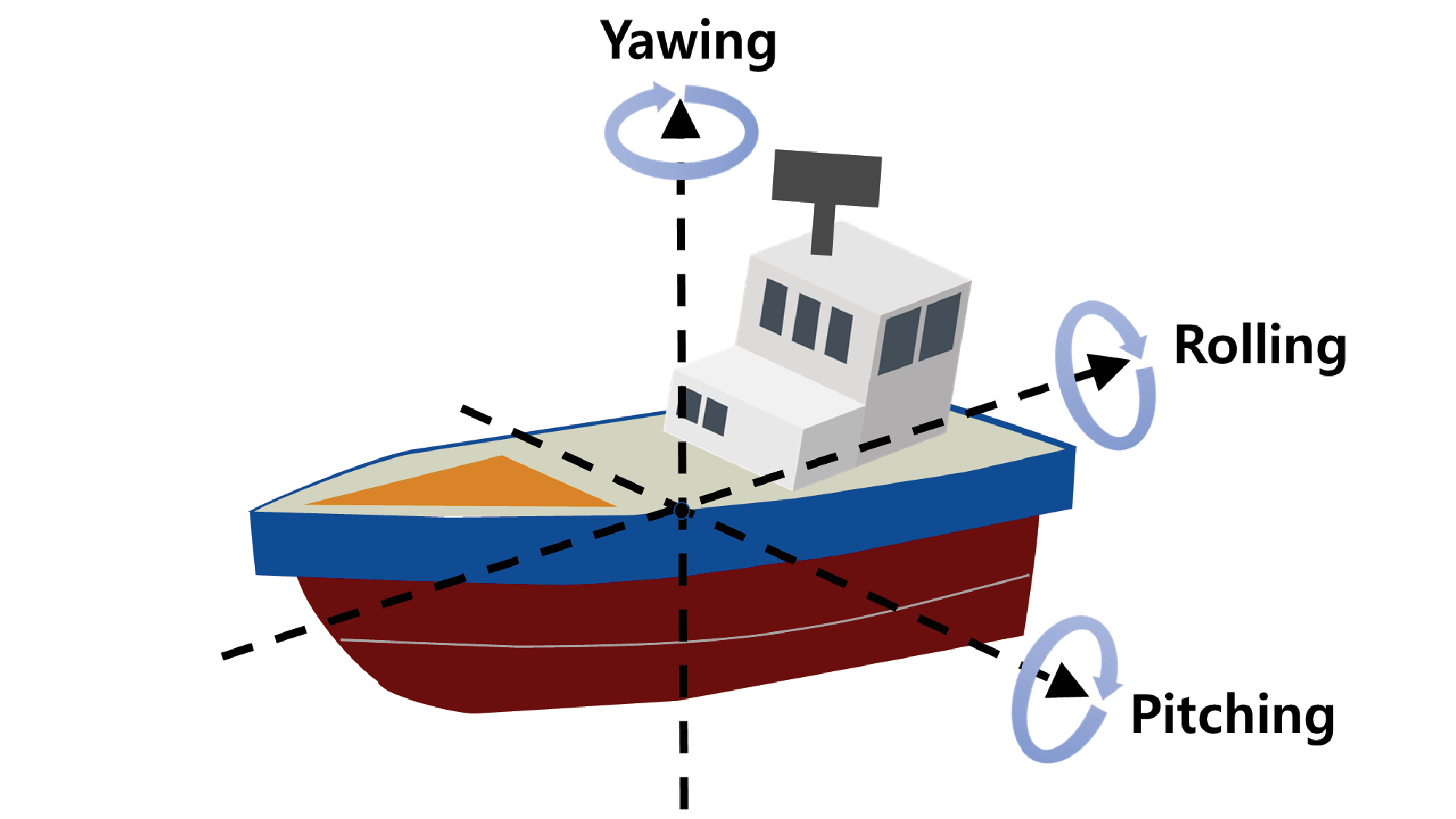}
  \caption{movemensts of the mobile platforms in maritime environment}
  \label{fig:mobile-platform}
\end{figure}

\subsection{Mobile Platform-based Pointing Error}
\label{pointing error}
In maritime environments, pointing errors are caused by the movements of mobile platforms, such as rolling, yawing, and pitching \cite{Kug2019}, as shown in Fig. \ref{fig:mobile-platform}.
Unlike conventional statistical model of pointing error in terrestrial environments, in which only the position changes due to building sway, both the position and orientation of the mobile platforms change in our system model.
Therefore, we applied the new statistical model of the pointing error derived in \cite{Marzieh} that considers the position, orientation, and non-orthogonality of the FSO beam.

Specifically, according to changes in the position and orientation of the mobile platforms, both are random vectors denoted by ${\bf{r}} = \left( {{r_x},{r_y},{r_z}} \right)$ and ${\boldsymbol{\omega }} = \left( {\theta ,\phi } \right)$.
Here, $\theta  \in \left[ {0,2\pi } \right]$ represents the angle between the projection of the beam vector onto the $x'-y'$ plane and $x'$ axis.
In addition, $\phi  \in \left[ {0,\pi } \right]$ denotes the angle between the beam vector and $z'$ axis.
We assume that all position and orientation variables are independent and follow Gaussian distribution, therefore,   ${\bf{r}}$ and ${\boldsymbol{\omega }}$ can be expressed as ${\bf{r}} = \left( {{\mu _x} + {\varepsilon _x}, {\mu _y} + {\varepsilon _y},{\mu _z} + {\varepsilon _z}} \right)$ and ${\boldsymbol{\omega }} = \left( {{\mu _\theta } + {\varepsilon _\theta },{\mu _\phi } + {\varepsilon _\phi }} \right)$, respectively.
Here, ${{\boldsymbol{\mu }}_{\bf{r}}} = \left( {{\mu _x},{\mu _y},{\mu _z}} \right)$ and ${{\boldsymbol{\mu }}_{\boldsymbol{\omega }}} = \left( {{\mu _\theta },{\mu _\phi }} \right)$ are means of ${\bf{r}}$ and ${\boldsymbol{\omega }}$, respectively and ${{\boldsymbol{\varepsilon }}_{\bf{r}}} = \left( {{\varepsilon _x},{\varepsilon _y},{\varepsilon _z}} \right)$ and ${{\boldsymbol{\varepsilon }}_{\boldsymbol{\omega }}} = \left( {{\varepsilon _\theta },{\varepsilon _\phi }} \right)$ are fluctuations of ${\bf{r}}$ and ${\boldsymbol{\omega }}$ modeled by zero-mean normal random variables with variance  $\sigma _i^2$, $i \in \left\{ {x,y,z,\theta ,\phi } \right\}$, respectively.

\begin{figure}[ht] % float placement: (h)ere, page (t)op, page (b)ottom, other (p)age
  \centering
  % file name: C:/Users/IBK/Desktop/Task2_converted_graphics/b.eps
  \includegraphics[width=13.0 cm]{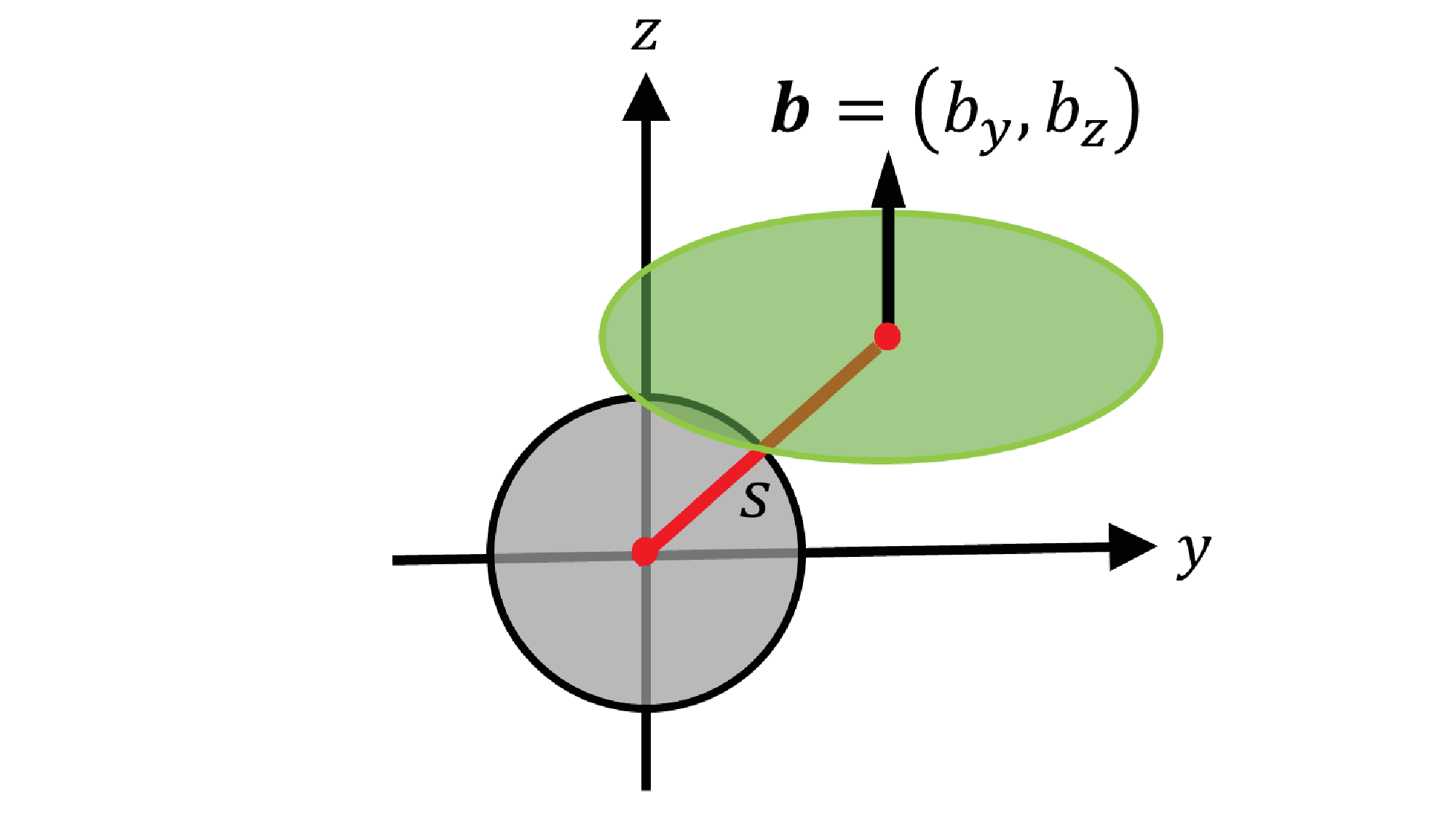}
  \caption{Beam footprint on the receiver lens plane}
  \label{fig:b}
\end{figure}

Based on \cite{Marzieh} and \cite{Kug2022}, we can approximated the fraction of power collected by the receiver lens as follows:
\begin{equation} \label{ip}
{I_p} \approx {A_0}\exp \left( { - \frac{{2{s^2}}}{{t{\omega _L}^2}}} \right).
\end{equation}
Here, the center of the footprint, ${\bf{b}}$, that changes due to these fluctuations can be obtained from the intersection of the laser beam and the $x$-plane of PD, which can be written as ${\bf{b}} = \left( {0,{r_y} - {r_x}\tan \theta ,{r_z} - {r_x}\frac{{\cot \phi }}{{\cos \theta }}} \right)$, as shown in Fig. \ref{fig:b}.
In addition, $s = \left\| {\bf{b}} \right\|$ is the distance between the center of the beam footprint and the center of the receiver lens, and $\omega _L$ is the beamwidth at distance $L$ (m).
Moreover, $t = \frac{{{t_1} + {t_2}}}{2}$, ${t_1} = \frac{{\sqrt \pi  {\mathop{\rm erf}\nolimits} \left( {{v_1}} \right)}}{{2{v_1}\exp \left( { - v_1^2} \right)}}$, ${t_2} = \frac{{\sqrt \pi  {\mathop{\rm erf}\nolimits} \left( {{v_2}} \right)}}{{2{v_2}\exp \left( { - v_2^2} \right){{\sin }^2}{\mu _\phi }{{\cos }^2}{\mu _\theta }}}$, ${v_1} = \frac{{{r_0}}}{{{w_L}}}\sqrt {\frac{\pi }{2}}$, and ${v_2} = {v_1}\left| {\sin \phi \cos \theta } \right|$ where ${\mathop{\rm erf}\nolimits} \left( x \right) = \frac{2}{{\sqrt \pi  }}\int_0^x {{e^{ - {t^2}}}dt}$ is the error function and $r_0$ is the radius of receiver lens and ${A_0} = {\mathop{\rm erf}\nolimits} \left( {{v_1}} \right){\mathop{\rm erf}\nolimits} \left( {{v_2}} \right)$ is the maximum fraction of optical power captured by the receiver lens at $s=0$.
Our pointing error model follows a Hoyt distribution $s \sim {\cal H}\left( {q,\Omega } \right)$ with $q = \sqrt {\frac{{{\rm{min}}\left\{ {{\lambda _1},{\lambda _2}} \right\}}}{{{\rm{max}}\left\{ {{\lambda _1},{\lambda _2}} \right\}}}}$ and $\Omega  = {\lambda _1} + {\lambda _2}$.
Here, ${\lambda _1}$ and ${\lambda _2}$ are the eigenvalues of the matrix, where $\sum {_{{\rm{IG}}}}$ is given by
\begin{equation}
\sum {_{{\rm{IG}}}}  = \left[ {\begin{array}{*{20}{c}}
{\sigma _y^2 + c_1^2\sigma _x^2 + c_2^2\sigma _\theta ^2}&{{c_1}{c_5}\sigma _x^2 + {c_2}{c_4}\sigma _\theta ^2}\\
{{c_1}{c_5}\sigma _x^2 + {c_2}{c_4}\sigma _\theta ^2}&{\sigma _z^2 + c_3^2\sigma _\phi ^2 + c_4^2\sigma _\theta ^2 + c_5^2\sigma _x^2}
\end{array}} \right],
\end{equation}
where ${c_1} =  - \tan {\mu _\theta }$, ${c_2} =  - \frac{{{\mu _x}}}{{{{\cos }^2}{\mu _\theta }}}$, ${c_3} =  - \frac{{{\mu _x}}}{{{{\sin }^2}{\mu _\phi }\cos {\mu _\theta }}}$, ${c_4} =  - \frac{{{\mu _x}\cot {\mu _\phi }\tan {\mu _\theta }}}{{\cos {\mu _\theta }}}$, and ${c_5} =  - \frac{{\cot {\mu _\phi }}}{{\cos {\mu _\theta }}}$.
Then, based on the distribution of $s$, the PDF of ${I_p}$ is given by \cite{Kug2022},
\begin{equation} \label{ippdf}
{f_{{I_p}}}({I_p}) = \frac{\xi }{{{A_0}}}{\left( {\frac{{{I_p}}}{{{A_0}}}} \right)^{\frac{{\left( {1 + {q^2}} \right)\xi }}{{2q}} - 1}}{I_0}\left( { - \frac{{\left( {1 - {q^2}} \right)\xi }}{{2q}}} \right)\ln \left( {\frac{{{I_p}}}{{{A_0}}}} \right),\quad 0 < {I_p} \le {A_0},
\end{equation}
where $\xi  = \frac{{\left( {1 + {q^2}} \right)tw_L^2}}{{4q\Omega }}$ is the jitter of the pointing error and ${I_0}\left(  \cdot  \right)$ denotes the zero-order modified Bessel function of the first kind \cite[Eq. (8.431.1)]{table}.

\section{Composite PDF}
{Based on our channel model assumptions, the composite PDF of irradiance $I$ can be written as follows:
\begin{equation}\label{pdfI}
{f_I}\left( I \right) 
=\int_{\frac{I}{{{A_0}}}}^1 {\frac{1}{{{I_a}}}{f_{{I_p}}}\left( {\frac{I}{{{I_a}}}} \right){f_{{I_a}}}\left( {{I_a}} \right)d{I_a}} .
\end{equation}
Using (\ref{pdfI}), as shown in Appendix \ref{composite PDF}, the closed-form expression of (\ref{pdfI}) can be obtained as follows: 
\begin{equation} 
\begin{aligned} \label{pdfI2}
{f_I}\left( I \right) &= \frac{{2{z^k}}}{{\Gamma \left( k \right)}}\frac{\xi }{{{A_0}}}{\left( {\frac{I}{{{A_0}}}} \right)^{z - 1}}\sum\limits_{m = 0}^\infty  {\frac{1}{{{{\left( {m!} \right)}^2}}}{{\left( { - \frac{{\left( {1 - {q^2}} \right)\xi }}{{2q}}} \right)}^{2m}}{{\left( {\ln \left( {\frac{{{A_0}}}{I}} \right)} \right)}^{k - 1}}  }\\
&\times{   \sum\limits_{n = 0}^\infty  \binom{k-1}{n} {{\left( { - \frac{1}{2}\ln \left( {\frac{{{A_0}}}{I}} \right)} \right)}^{ - n}}{{\left( { - \varpi } \right)}^{ - (2m + n) - 1}}\gamma \left( {2m + n + 1, - \frac{\varpi }{2}\ln \left( {\frac{{{A_0}}}{I}} \right)} \right)} .
\end{aligned}
\end{equation}
Note that, in (\ref{pdfI2}), using Stirling's approximation, the value of (\ref{pdfI2}) decays exponentially as the indices $m$ and $n$ increase \cite{Abramowitz}. Thus, the approximation by the truncation of index $n$ is close to the exact value, and an infinite summation in terms of index $m$ can be achieved with a finite number $M$.

In the following subsections, using (\ref{pdfI2}), we analyze the statistical characteristics according to the applied detection technique (i.e., HD or IM/DD techniques). According to the applied detection technique, the instantaneous SNR can be defined as $\gamma  = \frac{{{\eta ^r}{I^r}}}{{{N_0}}}$ where $r$ denotes the detection techniques (i.e., $r=1$ for HD and $r=2$ for IM/DD) and the related average SNR can be written as follows:
\begin{equation}
\mu  = \frac{{{\eta ^r}{{\left( {{\mathbb {E}}\left[ I \right]} \right)}^r}}}{{{N_0}}} = \frac{{{\eta ^r}{{\left( {{\mathbb {E}}\left[ {{I_a}} \right]} \right)}^r}{{\left( {{\mathbb {E}}\left[ {{I_p}} \right]} \right)}^r}}}{{{N_0}}},
\end{equation}
where ${\mathbb {E}}\left[ {{I_a}} \right] = {\left( {\frac{z}{{z + 1}}} \right)^k}$, ${\mathbb {E}}\left[ {{I_p}} \right] = \frac{{{A_0}\xi }}{{\sqrt {\left( {1 + q\xi } \right)\left( {1 + \frac{\xi }{q}} \right)} }}$, and  $\eta$ is the effective photoelectric conversion ratio.
Subsequently, depending on the applied detection technique, the PDF expression of the instantaneous SNR, $\gamma$, can be obtained as ${f_\gamma }\left( \gamma  \right) = {f_I}\left( {I\left( \gamma  \right)} \right)\left| {\frac{{dI}}{{d\gamma }}} \right|$.
Detailed analysis results according to the detection techniques are described in the following subsections.

\subsection{Closed-Form Expression of Composite PDF based on HD technique}
For the HD technique, with the relationship between the irradiance, $I$,  and the SNR, $\gamma$, $I$ can be defined as follows:
\begin{equation} \label{HDI}
I = \frac{{\gamma {N_0}}}{\eta } = \left\{ \begin{array}{l}
\frac{{{{\left( {\frac{z}{{z + 1}}} \right)}^k}\frac{{{A_0}\gamma \xi }}{{\sqrt {\left( {1 + q\xi } \right)\left( {1 + \frac{\xi }{q}} \right)} }}}}{{{\mu _{HD}}}}\\
\frac{{{{\left( {\frac{z}{{z + 1}}} \right)}^k}{A_0}\gamma }}{{{\mu _{HD}}}}\qquad\qquad{\rm{ for}}\;\;\xi^2 \; \gg\; {\rm{1   }}
\end{array} \right.,
\end{equation}
and
\begin{equation} \label{HDI-2}
\frac{{dI}}{{d\gamma }} = \left\{ \begin{array}{l}
\frac{{{{\left( {\frac{z}{{z + 1}}} \right)}^k}\frac{{{A_0}\xi }}{{\sqrt {\left( {1 + q\xi } \right)\left( {1 + \frac{\xi }{q}} \right)} }}}}{{{\mu _{HD}}}}\\
\frac{{{{\left( {\frac{z}{{z + 1}}} \right)}^k}{A_0}}}{{{\mu _{HD}}}}\qquad\qquad{\rm{ for}}\;\;\xi^2 \; \gg \;{\rm{1             }}
\end{array} \right.,
\end{equation}
where, ${\mu _{HD}} = \left\{ \begin{array}{l}
\frac{{{{\left( {\frac{z}{{z + 1}}} \right)}^k}\frac{{\eta {A_0}\xi }}{{\sqrt {\left( {1 + q\xi } \right)\left( {1 + \frac{\xi }{q}} \right)} }}}}{{{N_0}}}\\
\frac{{{{\left( {\frac{z}{{z + 1}}} \right)}^k}\eta {A_0}}}{{{N_0}}}  \qquad\qquad{\rm{for}}\;\;\xi^2  \;\gg \;{\rm{1}}
\end{array} \right.$ is the average SNR. 
From (\ref{HDI}), the SNR can be written as follows:
\begin{equation} \label{SNRHD}
\gamma  = \left\{ \begin{array}{l}
\frac{{{\mu _{HD}}I}}{{{{\left( {\frac{z}{{z + 1}}} \right)}^k}\frac{{{A_0}\xi }}{{\sqrt {\left( {1 + q\xi } \right)\left( {1 + \frac{\xi }{q}} \right)} }}}}\\
\frac{{{\mu _{HD}}I}}{{{{\left( {\frac{z}{{z + 1}}} \right)}^k}{A_0}}} \qquad\qquad{\rm{ for}}\;\;\xi^2\;  \gg\; {\rm{1             }}
\end{array} \right..
\end{equation}
Then, by replacing $I$ in (\ref{pdfI2}) with (\ref{HDI}) and applying the transformation of random variables (RVs), PDF of the SNR using the HD technique can be written as follows:
\begin{equation}
\begin{aligned} \label{pdfsnrHD}
&{f_\gamma }\left( \gamma  \right) = \frac{{2{z^k}}}{{\Gamma \left( k \right)}}\xi {\left( {\frac{{{{\left( {\frac{z}{{z + 1}}} \right)}^k}\frac{\xi }{{\sqrt {\left( {1 + q\xi } \right)\left( {1 + \frac{\xi }{q}} \right)} }}}}{{{\mu _{HD}}}}} \right)^z}\sum\limits_{m = 0}^\infty  {\frac{1}{{{{\left( {m!} \right)}^2}}}{{\left( { - \frac{{\left( {1 - {q^2}} \right)\xi }}{{2q}}} \right)}^{2m}}\sum\limits_{n = 0}^\infty  {\binom{k-1}{n}{{\left( { - \varpi } \right)}^{ - \left( {2m + n} \right) - 1}}} } \\
&\times {\left( { - 2} \right)^n}{\gamma ^{z - 1}}{\left( {\ln \left( {\frac{{{\mu _{HD}}}}{{\gamma {{\left( {\frac{z}{{z + 1}}} \right)}^k}\frac{\xi }{{\sqrt {\left( {1 + q\xi } \right)\left( {1 + \frac{\xi }{q}} \right)} }}}}} \right)} \right)^{k - 1 - n}}\gamma \left( {2m + n + 1, - \frac{\varpi }{2}\ln \left( {\frac{{{\mu _{HD}}}}{{\gamma {{\left( {\frac{z}{{z + 1}}} \right)}^k}\frac{\xi }{{\sqrt {\left( {1 + q\xi } \right)\left( {1 + \frac{\xi }{q}} \right)} }}}}} \right)} \right).
\end{aligned}
\end{equation}
From (\ref{pdfsnrHD}), as shown in Appendix \ref{AppenoutageprobabilityHD}, the outage probability with HD technique can be obtained as follows:
\begin{equation}\label{outage probabilityHDfinal}
\begin{aligned}
{F_\gamma }\left( x  \right) &= \int_0^x {{f_\gamma }\left( \gamma  \right)d\gamma } \\
&=\frac{{2{z^k}}}{{\Gamma \left( k \right)}}\xi {\sum\limits_{m = 0}^\infty  {\frac{1}{{{{\left( {m!} \right)}^2}}}\left( { - \frac{{\left( {1 - {q^2}} \right)\xi }}{{2q}}} \right)} ^{2m}}\sum\limits_{n = 0}^\infty \binom{k-1}{n} {{{\left( { - \varpi } \right)}^{ - \left( {2m + n} \right) - 1}}{{\left( { - 2} \right)}^n}}  \\
&\times \Bigg( 
\frac{{{{\left( { - \frac{\varpi }{2}} \right)}^{2m + n + 1}}\Gamma \left( {k + 2m + 1} \right)}}{{\left( {2m + n + 1} \right){{\left( { - \frac{\varpi }{2} + z} \right)}^{k + 2m + 1}}}}\times{}_2{F_1}\left( {1,k + 2m + 1;2m + n + 2;\frac{{ - \varpi }}{{ - \varpi  + 2z}}} \right)\\
&-
\sum\limits_{i = 0}^{k - 1 - n} {\frac{\left( {k - 1 - n} \right)!}{i!{z^{k - n - i }}}}\Bigg(  
  {{\left( { - \frac{\varpi }{2}} \right)}^{2m + n + 1}} {{\left( { - \frac{\varpi }{2} + z} \right)}^{ - (2m + n + 1) - i}}  \\
  & \times  
\Bigg(   
   {\gamma \bigg( {2m + n + 1 + i,\left( { - \frac{\varpi }{2} + z} \right)\ln \bigg( {\frac{{{\mu _{HD}}}}{{x{{\left( {\frac{z}{{z + 1}}} \right)}^k}\frac{\xi }{{\sqrt {\left( {1 + q\xi } \right)\left( {1 + \frac{\xi }{q}} \right)} }}}}} \bigg)} \bigg)}  \\
  &{ - 
\gamma \bigg( {2m + n + 1 , - \frac{\varpi }{2}\ln \bigg( {\frac{{{\mu _{HD}}}}{{x{{\left( {\frac{z}{{z + 1}}} \right)}^k}\frac{\xi }{{\sqrt {\left( {1 + q\xi } \right)\left( {1 + \frac{\xi }{q}} \right)} }}}}} \bigg)} \bigg){{\bigg( {\frac{{{\mu _{HD}}}}{{x{{\left( {\frac{z}{{z + 1}}} \right)}^k}\frac{\xi }{{\sqrt {\left( {1 + q\xi } \right)\left( {1 + \frac{\xi }{q}} \right)} }}}}} \bigg)}^{ - z}}} \\
  & \times {{{\bigg( {\ln \bigg( {\frac{{{\mu _{HD}}}}{{x{{\left( {\frac{z}{{z + 1}}} \right)}^k}\frac{\xi }{{\sqrt {\left( {1 + q\xi } \right)\left( {1 + \frac{\xi }{q}} \right)} }}}}} \bigg)} \bigg)}^i}}
  \Bigg) \Bigg) 
\Bigg).
\end{aligned}
\end{equation}

\subsection{Closed-Form Expression of Composite PDF based on IM/DD technique}
For the IM/DD technique, using the relationship between $I$ and $\gamma$, $I$ can be defined as follows:
\begin{equation}\label{IMDDI}
I = \frac{{\sqrt {\gamma {N_0}} }}{\eta } = \left\{ \begin{array}{l}
{\left( {\frac{z}{{z + 1}}} \right)^k}\frac{{{A_0}\xi }}{{\sqrt {\left( {1 + q\xi } \right)\left( {1 + \frac{\xi }{q}} \right)} }}\sqrt {\frac{\gamma }{{{\mu _{IM/DD}}}}} \\
{\left( {\frac{z}{{z + 1}}} \right)^k}{A_0}\sqrt {\frac{\gamma }{{{\mu _{IM/DD}}}}} \qquad\qquad{\rm{ for}}\;\;\xi^2 \; \gg \;{\rm{1             }}
\end{array} \right.,
\end{equation}
and
\begin{equation}
\frac{{dI}}{{d\gamma }} = \left\{ \begin{array}{l}
{\left( {\frac{z}{{z + 1}}} \right)^k}\frac{{{A_0}\xi }}{{\sqrt {\left( {1 + q\xi } \right)\left( {1 + \frac{\xi }{q}} \right)} }}\frac{1}{{2{\mu _{IM/DD}}\sqrt {\frac{\gamma }{{{\mu _{IM/DD}}}}} }}\\
{\left( {\frac{z}{{z + 1}}} \right)^k}{A_0}\frac{1}{{2{\mu _{IM/DD}}\sqrt {\frac{\gamma }{{{\mu _{IM/DD}}}}} }} \qquad\qquad{\rm{ for}}\;\;\xi^2  \;\gg \;{\rm{1              }}
\end{array} \right.,
\end{equation}
where, ${\mu _{IM/DD}} = \left\{ \begin{array}{l}
\frac{{{\eta ^2}{{\left( {{\mathbb E}\left[ I \right]} \right)}^2}}}{{{N_0}}} = {\textstyle{{{{\left( {{{\left( {\frac{z}{{z + 1}}} \right)}^k}\frac{{\eta {A_0}\xi }}{{\sqrt {\left( {1 + q\xi } \right)\left( {1 + \frac{\xi }{q}} \right)} }}} \right)}^2}} \over {{N_0}}}}\\
\frac{{{{\left( {{{\left( {\frac{z}{{z + 1}}} \right)}^k}\eta {A_0}} \right)}^2}}}{{{N_0}}} \qquad\qquad{\rm{for}}\;\;\xi^2  \;\gg\; {\rm{1}}
\end{array} \right.$ is the average SNR. 
Then, similar to the HD case, $\gamma$ can be written as follows:
\begin{equation}\label{SNRIMDD}
\gamma  = \left\{ \begin{array}{l}
{\mu _{IM/DD}}{\left( {{\textstyle{I \over {{{\left( {\frac{z}{{z + 1}}} \right)}^k}\frac{{{A_0}\xi }}{{\sqrt {\left( {1 + q\xi } \right)\left( {1 + \frac{\xi }{q}} \right)} }}}}}} \right)^2}\\
{\mu _{IM/DD}}{\left( {{\textstyle{I \over {{A_0}{{\left( {\frac{z}{{z + 1}}} \right)}^k}}}}} \right)^2} \quad\quad{\rm{ for}}\;\;\xi^2  \;\gg \;{\rm{1              }}
\end{array} \right..
\end{equation}
Then, by replacing $I$ in (\ref{pdfI2}) with (\ref{IMDDI}) and applying the transformation of the RVs, PDF of the SNR with IM/DD technique can be written as follows:
\begin{equation}
\begin{aligned} \label{pdfIMDD}
{f_\gamma }\left( \gamma  \right) &= \frac{{{z^k}}}{{\Gamma \left( k \right)}}\xi {\left( {{{\left( {\frac{z}{{z + 1}}} \right)}^k}\frac{\xi }{{\sqrt {\left( {1 + q\xi } \right)\left( {1 + \frac{\xi }{q}} \right)} }}} \right)^z}{\left( {\frac{1}{\mu_{IM/DD} }} \right)^{\frac{z}{2}}}\sum\limits_{m = 0}^\infty  {\frac{1}{{{{\left( {m!} \right)}^2}}}{{\left( { - \frac{{\left( {1 - {q^2}} \right)\xi }}{{2q}}} \right)}^{2m}}       }\\
&{ \times
\sum\limits_{n = 0}^\infty  {\binom{k-1}{n}{{\left( { - \varpi } \right)}^{ - \left( {2m + n} \right) - 1}}} }{\left( { - 2} \right)^n}{\gamma ^{\frac{z}{2} - 1}}      
{\left( {\ln \left( {\frac{{\sqrt {\frac{{{\mu _{IM/DD}}}}{\gamma }} }}{{{{\left( {\frac{z}{{z + 1}}} \right)}^k}\frac{\xi }{{\sqrt {\left( {1 + q\xi } \right)\left( {1 + \frac{\xi }{q}} \right)} }}}}} \right)} \right)^{k - 1 - n}}\\
&\times \gamma \left( {2m + n + 1, - \frac{\varpi }{2}\ln \left( {\frac{{\sqrt {\frac{{{\mu _{IM/DD}}}}{\gamma }} }}{{{{\left( {\frac{z}{{z + 1}}} \right)}^k}\frac{\xi }{{\sqrt {\left( {1 + q\xi } \right)\left( {1 + \frac{\xi }{q}} \right)} }}}}} \right)} \right).
\end{aligned}
\end{equation}
From (\ref{pdfIMDD}), as shown in Appendix \ref{AppenoutageprobabilityIMDD}, the outage probability with IM/DD technique can be obtained as follows:
\begin{equation} \label{outage probabilityIMDDfinal}
\begin{aligned}
{F_\gamma }\left( x  \right) &= \frac{{2{z^k}}}{{\Gamma \left( k \right)}}\xi {\sum\limits_{m = 0}^\infty  {\frac{1}{{{{\left( {m!} \right)}^2}}}\left( { - \frac{{\left( {1 - {q^2}} \right)\xi }}{{2q}}} \right)} ^{2m}}\sum\limits_{n = 0}^\infty \binom{k-1}{n} {{{\left( { - \varpi } \right)}^{ - \left( {2m + n} \right) - 1}}{{\left( { - 2} \right)}^n}}  \\
&\times \Bigg( 
\frac{{{{\left( { - \frac{\varpi }{2}} \right)}^{2m + n + 1}}\Gamma \left( {k + 2m + 1} \right)}}{{\left( {2m + n + 1} \right){{\left( { - \frac{\varpi }{2} + z} \right)}^{k + 2m + 1}}}}\times{}_2{F_1}\left( {1,k + 2m + 1;2m + n + 2;\frac{{ - \varpi }}{{ - \varpi  + 2z}}} \right)\\
&-
\sum\limits_{i = 0}^{k - 1 - n} {\frac{\left( {k - 1 - n} \right)!}{i!{z^{k - n - i }}}}\Bigg(  
  {{\left( { - \frac{\varpi }{2}} \right)}^{2m + n + 1}} {{\left( { - \frac{\varpi }{2} + z} \right)}^{ - (2m + n + 1) - i}}  \\
  & \times  \Bigg(   
   {\gamma \bigg( {2m + n + 1 + i,\left( { - \frac{\varpi }{2} + z} \right)\ln \bigg( {\frac{{\sqrt {\frac{{{\mu _{IM/DD}}}}{x}} }}{{{{\left( {\frac{z}{{z + 1}}} \right)}^k}\frac{\xi }{{\sqrt {\left( {1 + q\xi } \right)\left( {1 + \frac{\xi }{q}} \right)} }}}}} \bigg)} \bigg)}  \\
  &{ -
 \gamma \bigg( {2m + n + 1 , - \frac{\varpi }{2}\ln \bigg( {\frac{{\sqrt {\frac{{{\mu _{IM/DD}}}}{x}} }}{{{{\left( {\frac{z}{{z + 1}}} \right)}^k}\frac{\xi }{{\sqrt {\left( {1 + q\xi } \right)\left( {1 + \frac{\xi }{q}} \right)} }}}}} \bigg)} \bigg){{\bigg( {\frac{{\sqrt {\frac{{{\mu _{IM/DD}}}}{x}} }}{{{{\left( {\frac{z}{{z + 1}}} \right)}^k}\frac{\xi }{{\sqrt {\left( {1 + q\xi } \right)\left( {1 + \frac{\xi }{q}} \right)} }}}}} \bigg)}^{ - z}}} \\
  & \times {{{\bigg( {\ln \bigg( {\frac{{\sqrt {\frac{{{\mu _{IM/DD}}}}{x}} }}{{{{\left( {\frac{z}{{z + 1}}} \right)}^k}\frac{\xi }{{\sqrt {\left( {1 + q\xi } \right)\left( {1 + \frac{\xi }{q}} \right)} }}}}} \bigg)} \bigg)}^i}}
  \Bigg) \Bigg) \Bigg).
\end{aligned}
\end{equation}

\subsection{Closed-Form expression of Unified Composite PDF}
In this subsection, the PDF results of SNR derived from HD and IM/DD techniques given in (\ref{pdfsnrHD}) and (\ref{pdfIMDD}), respectively, in the previous subsections are rewritten in a unified form expression as follows:
\begin{equation}
\begin{aligned} \label{unifiedpdf}
{f_\gamma }\left( \gamma  \right) 
&= \frac{{2{z^k}}}{{\Gamma \left( k \right)}}\frac{\xi}{r} {\left( {{{\left( {\frac{z}{{z + 1}}} \right)}^k}\frac{\xi }{{\sqrt {\left( {1 + q\xi } \right)\left( {1 + \frac{\xi }{q}} \right)} }}} \right)^z}{\left( {\frac{1}{\mu }} \right)^{\frac{z}{r}}}\sum\limits_{m = 0}^\infty  {\frac{1}{{{{\left( {m!} \right)}^2}}}{{\left( { - \frac{{\left( {1 - {q^2}} \right)\xi }}{{2q}}} \right)}^{2m}}       }\\
&{ \times
\sum\limits_{n = 0}^\infty  {\binom{k-1}{n}{{\left( { - \varpi } \right)}^{ - \left( {2m + n} \right) - 1}}} }{\left( { - 2} \right)^n}{\gamma ^{\frac{z}{r} - 1}}      
{\left( {\ln \left( {\frac{{\left( {\frac{{{\mu}}}{\gamma }}\right)^{{{\frac{1}{r}}}} }}{{{{\left( {\frac{z}{{z + 1}}} \right)}^k}\frac{\xi }{{\sqrt {\left( {1 + q\xi } \right)\left( {1 + \frac{\xi }{q}} \right)} }}}}} \right)} \right)^{k - 1 - n}}\\
&\times {\gamma \left( {2m + n + 1, - \frac{\varpi }{2}\ln \left( {\frac{{{{\left( {\frac{\mu }{\gamma }} \right)}^{\frac{1}{r}}}}}{{{{\left( {\frac{z}{{z + 1}}} \right)}^k}\frac{\xi }{{\sqrt {\left( {1 + q\xi } \right)\left( {1 + \frac{\xi }{q}} \right)} }}}}} \right)} \right)},
\end{aligned}
\end{equation}
where $r=1$ with HD and $r=2$ with IM/DD techniques. 
Using Stirling's approximation (as $m$ and $n$ increase, infinite summations decrease exponentially), infinite summations can be achieved using a finite number of terms.

For ${\xi ^2} \gg 1$, it can be assumed that the pointing error is negligible. Therefore, the irradiance can be simplified to $I \approx {I_a}$.
Therefore, the PDF of $I$ can be expressed as follows:
\begin{equation}
{f_I}\left( I \right) = \frac{{{z^k}}}{{\Gamma (k)}}{\left[ {\ln \left( {\frac{1}{I}} \right)} \right]^{k - 1}}{I^{z - 1}}.
\end{equation}
Then, by replacing $I$ with the unified SNR representation given in (\ref{SNRHD}) and (\ref{SNRIMDD}) and applying the transformation of RVs, the PDF of SNR in a unified form expression can be written as follows:
\begin{equation} \label{infPDF}
{f_\gamma }\left( \gamma  \right) = \frac{{{z^k}}}{{\Gamma (k)}}{{{{{\left( {{A_o}{{\left( {\frac{z}{{z + 1}}} \right)}^k}{{\left( {\frac{\gamma }{\mu }} \right)}^{\frac{1}{r}}}} \right)}^z}} \over {{\gamma ^2}}}}{\left[ {\ln \left( {{\textstyle{1 \over {{A_o}{{\left( {\frac{z}{{z + 1}}} \right)}^k}{{\left( {\frac{\gamma }{\mu }} \right)}^{\frac{1}{r}}}}}}} \right)} \right]^{k - 1}}.
\end{equation}

\subsection{Closed-Form expression of Unified Composite PDF based on TMOS}
Based on the system and channel models, the range of the optical beam (i.e., irradiance) affected by the foggy channel and pointing error is $0 < I \le {A_0}$. Then, according to the relationship between $I$ and $\gamma$, the range of $\gamma$ can be obtained as $0 < \gamma \le \frac{\mu }{{{{\left( {{{\left( {\frac{z}{{z + 1}}} \right)}^k}\frac{\xi }{{\sqrt {\left( {1 + q\xi } \right)\left( {1 + \frac{\xi }{q}} \right)} }}} \right)}^r}}}$. In addition, based on the TMOS scheme, the SNR of the selected beam is above the preselected threshold $\gamma_{T}$. Therefore, based on \cite{Sung2017}, the statistical representations of the output SNR of the selected beam can finally be obtained as follows:
\begin{equation}\label{statisticalSNR}
{f_{TB\_WDM}}\left( \gamma  \right) = \left\{ \begin{array}{lr}
{{{{f_\gamma }\left( \gamma  \right)} \over {{F_\gamma }\left( {\frac{\mu }{{{{\left( {{{\left( {\frac{z}{{z + 1}}} \right)}^k}\frac{\xi }{{\sqrt {\left( {1 + q\xi } \right)\left( {1 + \frac{\xi }{q}} \right)} }}} \right)}^r}}}} \right) - {F_\gamma } 
\left( {{\gamma _T}} \right)}}} \quad {\gamma _T} \le \gamma  \le \frac{\mu }{{{{\left( {{{\left( {\frac{z}{{z + 1}}} \right)}^k}\frac{\xi }{{\sqrt {\left( {1 + q\xi } \right)\left( {1 + \frac{\xi}{q}} \right)} }}} \right)}^r}}}\\
0 {\rm{ \qquad\qquad\qquad\qquad\qquad\qquad\qquad\quad otherwise}}
\end{array} \right.,
\end{equation}
where ${f_\gamma }\left( \gamma  \right)$ and ${F_\gamma }\left(  \cdot  \right)$ are the PDF and the outage probability of the SNR, respectively.

\section{Performance analysis of the Outage Probability}
Based on the operation mode, the outage probability, ${{\mathop{\rm P}\nolimits} _{OUT}}$, (i.e., equal to the probability of no transmission) is defined as the probability 
that the output (electrical) SNR of the selected beam with TMOS falls below a preset threshold.
Therefore, ${{\mathop{\rm P}\nolimits} _{OUT}}$ can be formulated as
\begin{equation}\label{POUT}
{{\mathop{\rm P}\nolimits} _{OUT}} = \Pr \left[ \gamma_{T} < {{\gamma _{TB\_WDM}} < {\gamma _{TH\_OUT}}} \right] = \int_{\gamma _T}^{{\gamma _{TH\_OUT}}} {{f_{TB\_WDM}}\left( \gamma  \right)d\gamma }.
\end{equation}
According to the TMOS scheme, the selected beam follows the conditional PDF of a truncated RV.
Subsequently, by substituting (\ref{statisticalSNR}) into (\ref{POUT}), a closed-form expression for ${{\mathop{\rm P}\nolimits} _{OUT}}$  can be obtained as follows:
\begin{equation} \label{expressionPOUT}
{{\mathop{\rm P}\nolimits} _{OUT}} = \frac{{{F_\gamma }\left( {{\gamma _{TH\_OUT}}} \right) - {F_\gamma }\left( {{\gamma _T}} \right)}}{{{F_\gamma }\left( {\frac{\mu }{{{{\left( {{{\left( {\frac{z}{{z + 1}}} \right)}^k}\frac{\xi }{{\sqrt {\left( {1 + q\xi } \right)\left( {1 + \frac{\xi }{q}} \right)} }}} \right)}^r}}}} \right) - {F_\gamma }\left( {{\gamma _T}} \right)}}= \frac{{{F_\gamma }\left( {{\gamma _{TH\_OUT}}} \right) - {F_\gamma }\left( {{\gamma _T}} \right)}}{{1 - {F_\gamma }\left( {{\gamma _T}} \right)}}.
\end{equation}
Subsequently, in (\ref{expressionPOUT}), the final closed-form results can be obtained by simply replacing ${F_\gamma }\left(  \cdot  \right)$ with (\ref{outage probabilityHDfinal}) for HD technique and (\ref{outage probabilityIMDDfinal}) for IM/DD technique.

\section{Performance analysis with Adaptive modulation}

\begin{table}[ht]
\centering
\caption{Parameters $a_u$ and $b_u$ based on the given $M_u$, ${\gamma}_{T_u}$ for Target $\overline {BE{R_0}}  = {10^{ - 3}}$ }
\label{AM}
\begin{tabular}{c|c|c|c|c}
\noalign{\smallskip}\noalign{\smallskip}\hline
$u$ & $M_u$ & $a_u$ & $b_u$ & ${\gamma}_{T_u}\;[dB]$ \\
\hline
1  & 4 & 896.0704 & 10.7367 & 7.1 \\

2  & 8 & 404.4353 & 6.8043 & 11.8 \\

3  & 16 & 996.5492 & 8.7345 & 14.0 \\

4  & 32 & 443.1272 & 8.2282 & 17.0 \\

5  & 64 & 296.6007 & 7.9270 & 20.1 \\

6  & 128 & 327.4874 & 8.2036 & 23.0 \\

7  & 256 & 404.2837 & 7.8824 & 26.2 \\

8  & 512 & 310.5283 & 8.2425 & 29.0 \\

\hline
\end{tabular}
\end{table}

\subsection{Average Spectral Efficiency }
{ Based on the system models described in Sec. II, the ASE of the selected beam can be estimated as the sum of all spectral efficiencies, $R_u$, weighted by the probability, $F_u$, that the SNR of a selected beam is assigned to the $u$-th region as follows:  
\begin{equation} \label{ASE}
ASE = \sum\limits_{u = 1}^{{N_{\max }}} {{R_u}{F_u}},
\end{equation}
where ${R_u} = u + 0.5$ and ${F_u}$ can be considered the following two cases depending on the corresponding integral region according to ${\textstyle{\mu  \over {{{\left( {{{\left( {\frac{z}{{z + 1}}} \right)}^k}\frac{\xi }{{\sqrt {\left( {1 + q\xi } \right)\left( {1 + \frac{\xi }{q}} \right)} }}} \right)}^r}}}}$.
Let ${\textstyle{\mu  \over {{{\left( {{{\left( {\frac{z}{{z + 1}}} \right)}^k}\frac{\xi }{{\sqrt {\left( {1 + q\xi } \right)\left( {1 + \frac{\xi }{q}} \right)} }}} \right)}^r}}}}$ be $\gamma_{max}$, then, by substituting ${f_{TB\_WDM}}\left(  \cdot  \right)$ with (\ref{statisticalSNR}), we can obtain the closed-form expression of ${F_u}$ equation required for estimating the ASE as follows:

Case 1. ${\gamma _{{T_N}}} \le \gamma_{max} < {\gamma _{{T_{N + 1}}}}, \;\;{N<N_{\max }}$,
\begin{equation} \label{ASE_1}
{F_u} = \left\{ \begin{array}{ll}
\int_{{\gamma _{{T_u}}}}^{{\gamma _{{T_{u + 1}}}}} {{f_{TB\_WDM}}\left( \gamma  \right)d\gamma }
%\\
= {{{{F_\gamma }\left( {{\gamma _{{T_{u + 1}}}}} \right) - {F_\gamma }\left( {{\gamma _{{T_u}}}} \right)} \over {{F_\gamma }\left( \gamma_{max} \right) - {F_\gamma }\left( {{\gamma _T}} \right)}}}  \quad\;\;\;&u = 1, \cdots ,N- 1
\\\\
\int_{{\gamma _{T_{N}}}}^{\gamma_{max}} {{f_{TB\_WDM}}\left( \gamma  \right)d\gamma }
%\\
 ={{{{F_\gamma }\left( \gamma_{max} \right) - {F_\gamma }\left( {{\gamma _{T_{N}}}} \right)} \over {{F_\gamma }\left( \gamma_{max} \right) - {F_\gamma }\left( {{\gamma _T}} \right)}}} \quad &u = N
\end{array} \right.,
\end{equation}

Case 2. $\gamma_{max} > {\gamma _{T_{N_{\max }}}}$,
\begin{equation} \label{ASE_2}
{F_u} = \left\{ \begin{array}{l}
\int_{{\gamma _{{T_u}}}}^{{\gamma _{{T_{u + 1}}}}} {{f_{TB\_WDM}}\left( \gamma  \right)d\gamma }
%\\
= {{{{F_\gamma }\left( {{\gamma _{{T_{u + 1}}}}} \right) - {F_\gamma }\left( {{\gamma _{{T_u}}}} \right)} \over {{F_\gamma }\left( \gamma_{max} \right) - {F_\gamma }\left( {{\gamma _T}} \right)}}}  \quad\quad\;\;u = 1, \cdots ,{N_{\max }} - 1
\\\\
\int_{\gamma _{T_{N_{\max }}}}^{\gamma_{max}} {{f_{TB\_WDM}}\left( \gamma  \right)d\gamma }
%\\
= {{{{F_\gamma }\left( {{\gamma _{\max }}} \right) - {F_\gamma }\left( {\gamma _{T_{N_{\max }}}} \right)} \over {{F_\gamma }\left( \gamma_{max} \right) - {F_\gamma }\left( {{\gamma _T}} \right)}}} \quad \quad\quad\;\; u = {N_{\max }}
\end{array} \right..
\end{equation}
We can then obtain the closed-form results of $F_{u}$ by replacing ${F_\gamma }\left(  \cdot  \right)$ in (\ref{ASE_1}) and (\ref{ASE_2}) with (\ref{outage probabilityHDfinal}) for HD technique and (\ref{outage probabilityIMDDfinal}) for IM/DD technique, respectively.
Finally, by substituting $F_{u}$ and $R_{u}$ in (\ref{ASE}) with the closed-form results above, we obtain the closed-form results of the ASE for each detection technique.

\subsection{Average Bit Error Rate }
In the case of ACM, we can express the average BER, $\overline {BER}$, of all the codes for the SNRs of the selected beam as the average number of bits in the error divided by the ASE in (\ref{ASE}) as follows:
\begin{equation}\label{AvgBER}
\overline {BER}  = \frac{{\sum\limits_{u = 1}^{{N_{\max }}} {{R_u}{{\overline {BER} }_u}} }}{{\sum\limits_{u = 1}^{{N_{\max }}} {{R_u}{F_u}} }}.
\end{equation}
Similar to the ASE analysis case, evaluating ${{\overline {BER} }_u}$ can be considered in the following two cases, depending on the corresponding integral region based on  ${\gamma_{max}}$.
In (\ref{AvgBER}), to obtain the closed-form results for $\overline {BER}$, we only need to derive the closed-form results for ${{{\overline {BER} }_u}}$.  Thus, in the following, we focus on the derivation of the closed-form results of unified ${{{\overline {BER} }_u}}$ for HD and IM/DD methods.

Case 1. ${\gamma _{{T_N}}} \le \gamma_{max} < {\gamma _{{T_{N + 1}}}}, N<{N_{\max }}$
\begin{equation} \label{AvgBER1}
{\overline {BER} _u} \!=\!\! \left\{ \begin{array}{l}
\!\!\!\!\int_{{\gamma _{{T_u}}}}^{{\gamma _{{T_{u + 1}}}}} \!\!{BE{R_u}{f_{TB\_WDM}}\!\left( \gamma  \right)d\gamma } \;\;\;u = 1, \cdots ,{N} - 1\\\\
\!\!\!\!\int_{{\gamma _{T_{N}}}}^ {\gamma_{max}} \!\!{BE{R_u}{f_{TB\_WDM}}\!\left( \gamma  \right)d\gamma } \;\;\;u = {N}
\end{array} \right.,
\end{equation}
\\

Case 2. $ {\gamma_{max}} > {\gamma _{T_{N_{\max }}}}$
\begin{equation} \label{AvgBER2} 
{\overline {BER} _u} \!=\!\! \left\{ \begin{array}{l}
\!\!\!\!\int_{{\gamma _{{T_u}}}}^{{\gamma _{{T_{u + 1}}}}} \!\!{BE{R_u}{f_{TB\_WDM}}\left( \gamma  \right)d\gamma } \;\;\;u = 1, \cdots ,{N_{\max }} - 1\\\\
\!\!\!\!\int_{{\gamma _{T_{N_{\max }}}}}^ {\gamma_{max}} \!\!{BE{R_u}{f_{TB\_WDM}}\!\left( \gamma  \right)d\gamma } \;\;\;\;\;u = {N_{\max }}
\end{array} \right.,
\end{equation}
where $BER{_u} = {a_u}\exp \left( { - \frac{{{b_u}\gamma }}{{{M_u}}}} \right)$ is the BER for code $u$ according to \cite{Kjell}. Furthermore, the corresponding values of $a_u$, $b_u$, and $M_u$ are listed in Table \ref{AM} and ${M_u} = {2^{u + 1}} , u \in \left\{ {1,2, \ldots ,8} \right\}$ is the constellation size.
Here, by applying Maclaurin series of exponential function to $BER{_u}$ (i.e. $\exp \left( { - \frac{{{b_u}\gamma }}{{{M_u}}}} \right) = \sum\limits_{l = 0}^\infty  {\frac{{{{\left( { - \frac{{{b_u}\gamma }}{{{M_u}}}} \right)}^l}}}{{l!}}}$), (\ref{AvgBER1}) and (\ref{AvgBER2}) can be rewritten as follows:
 }

{
\begin{equation} \label{AvgBERu}
\begin{aligned} 
\!{\overline {BER} _u} \!\!&=\!\! \int_{{\gamma _{{T_u}}}}^{ {\gamma_{max}}}\!\!\!\!\!\! {BE{R_u}{f_{TB\_WDM}}\!\left(\! \gamma  \right)\!d\gamma } \!\! - \!\!\!\int_{{\gamma _{{T_{u + 1}}}}}^{ {\gamma_{max}}}\!\!\!\!\!\! {BE{R_u}{f_{TB\_WDM}}\!\left(\! \gamma  \right)\!d\gamma } + {\int_{{\gamma _{{T_{{N_{\max }}}}}}}^{{\gamma _{\max }}} {BE{R_u}{f_{TB\_WDM}}\left( \gamma  \right)d\gamma }  }   \\
&= {\overline {BER} _{u,{\rm I}}} - {\overline {BER} _{u,{\rm I}{\rm I}}} + {\overline {BER} _{u,{\rm I}{\rm I}{\rm I}}},
\end{aligned}
\end{equation}
where,
\begin{equation} \label{BERuI}
{\overline {BER} _{u,{\rm I}}}\! =\! \sum\limits_{l = 0}^\infty  {\frac{{{a_u}{{\left( { - \frac{{{b_u}}}{{{M_u}}}} \right)}^l}}}{{l!}} \frac{1}{{{F_\gamma }\left(  { {\gamma_{max}}} \right)\! -\! {F_\gamma }\left( {{\gamma _T}} \right)}}} \int_{{\gamma _{{T_u}}}}^{ {\gamma_{max}}}\!\!\!\!{{\gamma ^l}} {f_\gamma }\left( \gamma  \right)d\gamma ,
\end{equation}
\begin{equation} \label{BERuII}
{\overline {BER} _{u,{\rm I}{\rm I}}} \!=\! \sum\limits_{l = 0}^\infty  {\frac{{{a_u}{{\left( { - \frac{{{b_u}}}{{{M_u}}}} \right)}^l}}}{{l!}} \frac{1}{{{F_\gamma }\left(  { {\gamma_{max}}} \right)\! -\! {F_\gamma }\left( {{\gamma _T}} \right)}}} \int_{{\gamma _{{T_{u + 1}}}}}^{ {\gamma_{max}}}\!\!\!\!{{\gamma ^l}} {f_\gamma }\left( \gamma  \right)d\gamma ,
\end{equation} 
and
\begin{equation} \label{BERuIII}
{{\overline {BER} _{u,{\rm I}{\rm I}{\rm I}}} = \sum\limits_{l = 0}^\infty  {\frac{{{a_n}{{\left( { - \frac{{{b_n}}}{{{M_n}}}} \right)}^l}}}{{l!}}  \frac{1}{{{F_\gamma }\left( {{\gamma _{\max }}} \right) - {F_\gamma }\left( {{\gamma _T}} \right)}}} \int_{{\gamma _{{T_{{N_{\max }}}}}}}^{{\gamma _{\max }}} {{\gamma ^l}} {f_\gamma }\left( \gamma  \right)d\gamma .}
\end{equation}
{ By replacing ${{\gamma _{{T_{{N_{\max }}}}}}}$ in (\ref{BERuIII}) with ${{\gamma _{{T_{{8}}}}}}$, we can express Case 2.}
Using (\ref{BERuI}), (\ref{BERuII}), and (\ref{BERuIII}), by substituting {
(\ref{pdfsnrHD}) and (\ref{outage probabilityHDfinal}) for HD technique and (\ref{pdfIMDD}) and (\ref{outage probabilityIMDDfinal}) for IM/DD technique into ${{f_\gamma }\left( \gamma  \right)}$ and ${F_\gamma }\left( x  \right)$,} we can derive the closed-form expressions of (\ref{BERuI}), (\ref{BERuII}), and (\ref{BERuIII}) for both HD and IM/DD techniques, respectively.

In addition, in (\ref{BERuI}), (\ref{BERuII}), and (\ref{BERuIII}), using Stirling's approximation (as $l$ increases, the infinite summation decreases exponentially), the required accuracy can be achieved by a truncated summation with a finite number of terms.
The detailed closed-form result derivation process of (\ref{BERuI}), (\ref{BERuII}), and (\ref{BERuIII}) with both HD and IM/DD techniques are presented in Appendix \ref{AppenBERHD} and \ref{AppenBERIMDD}, respectively.
With these closed-form results for both HD and IM/DD techniques, the unified closed-form result expression of ${\overline {BER} _{u,{\rm I}}}$ can be expressed as follows:
\begin{equation}  \label{unifiedBERuI}
\begin{aligned}
  {\overline {BER} _{u,{\rm I}}} &= \sum\limits_{l = 0}^\infty  {\frac{{{a_u}{{\left( { - \frac{{{b_u}}}{{{M_u}}}} \right)}^l}}}{{l!}}}{{\textstyle{1 \over {{F_\gamma }\left( {\frac{{{\mu }}}{{{{\left( {{{\left( {\frac{z}{{z + 1}}} \right)}^k}\frac{\xi }{{\sqrt {\left( {1 + q\xi } \right)\left( {1 + \frac{\xi }{q}} \right)} }}} \right)}^r}}}} \right) - {F_\gamma }\left( {{\gamma _T}} \right)}}}} \frac{{2{z^k}}}{{\Gamma \left( k \right)}}\xi {\left( {{\textstyle{\mu  \over {{{\left( {{{\left( {\frac{z}{{z + 1}}} \right)}^k}\frac{\xi }{{\sqrt {\left( {1 + q\xi } \right)\left( {1 + \frac{\xi }{q}} \right)} }}} \right)}^r}}}}} \right)^{l}} \\
  &\times 
{\sum\limits_{m = 0}^\infty  {\frac{1}{{{{\left( {m!} \right)}^2}}}\left( { - \frac{{\left( {1 - {q^2}} \right)\xi }}{{2q}}} \right)} ^{2m}}\sum\limits_{n = 0}^\infty  \binom{k-1}{n} {\left( { - \varpi } \right)^{ - \left( {2m + n} \right) - 1}}{\left( { - 2} \right)^n}\\
  &\times
\sum\limits_{i = 0}^{k - 1 - n}{\frac{{\left( {k - 1 - n} \right)!}}{{i!{{\left( {rl + z} \right)}^{k - n - i}}}}} 
\Bigg( \Bigg. 
  {{\left( { - \frac{\varpi }{2}} \right)}^{2m + n + 1}} {{\left( { - \frac{\varpi }{2} +rl+ z} \right)}^{ - (2m + n + 1) - i}}  \\
  & \times    
   {\gamma \bigg( {2m + n + 1 + i,\left( { - \frac{\varpi }{2} +rl+ z} \right)\ln \bigg( {\frac{{ \left({\frac{{{\mu }}}{{{\gamma _{{T_u}}}}}}\right)^{\frac{1}{r}} }}{{{{\left( {\frac{z}{{z + 1}}} \right)}^k}\frac{\xi }{{\sqrt {\left( {1 + q\xi } \right)\left( {1 + \frac{\xi }{q}} \right)} }}}}} \bigg)} \bigg)}  \\
  &{ - \gamma \bigg( {2m + n + 1 , - \frac{\varpi }{2}\ln \bigg( {\frac{{ \left({\frac{{{\mu }}}{{{\gamma _{{T_u}}}}}}\right)^{\frac{1}{r}} }}{{{{\left( {\frac{z}{{z + 1}}} \right)}^k}\frac{\xi }{{\sqrt {\left( {1 + q\xi } \right)\left( {1 + \frac{\xi }{q}} \right)} }}}}} \bigg)} \bigg)
{{\bigg({\frac{{ \left({\frac{{{\mu }}}{{{\gamma _{{T_u}}}}}}\right)^{\frac{1}{r}} }}{{{{\left( {\frac{z}{{z + 1}}} \right)}^k}\frac{\xi }{{\sqrt {\left( {1 + q\xi } \right)\left( {1 + \frac{\xi }{q}} \right)} }}}}} \bigg)}^{{ - \left( {rl + z} \right)}}}} \\
  & \times {{{\bigg( {\ln \bigg({\frac{{ \left({\frac{{{\mu }}}{{{\gamma _{{T_u}}}}}}\right)^{\frac{1}{r}} }}{{{{\left( {\frac{z}{{z + 1}}} \right)}^k}\frac{\xi }{{\sqrt {\left( {1 + q\xi } \right)\left( {1 + \frac{\xi }{q}} \right)} }}}}} \bigg)} \bigg)}^i}}
  \Bigg).
\end{aligned}
\end{equation}

For ${\xi ^2} \gg 1$, by substituting (\ref{BERuI}) into (\ref{infPDF}), the unified expression for ${\overline {BER} _{u,{\rm I}}}$ can be expressed as
\begin{equation} \label{BERuI1}
\begin{aligned}
{{\overline {BER} _{u,{\rm I}}} }&= \sum\limits_{l = 0}^\infty  {\frac{{{a_u}{{\left( { - \frac{{{b_u}}}{{{M_u}}}} \right)}^l}}}{{l!}} \cdot {\textstyle{1 \over {{F_\gamma }\left( {{\textstyle{\mu  \over {{{\left( {{A_o}{{\left( {\frac{z}{{z + 1}}} \right)}^k}} \right)}^r}}}}} \right) - {F_\gamma }\left( {{\gamma _T}} \right)}}}} \\
& \times 
\int_{{\gamma _{{T_u}}}}^{{\textstyle{\mu  \over {{{\left( {{A_o}{{\left( {\frac{z}{{z + 1}}} \right)}^k}} \right)}^r}}}}} {{}\frac{{{z^k}\gamma ^{l-2}}}{{\Gamma (k)}}{{\Bigg( {\ln \bigg( {\frac{1}{{{A_o}{{\left( {\frac{z}{{z + 1}}} \right)}^k}{{\left( {\frac{\gamma }{\mu }} \right)}^{\frac{1}{r}}}}}} \bigg)} \Bigg)}^{k - 1}}{{{{\left( {{A_o}{{\left( {\frac{z}{{z + 1}}} \right)}^k}{{\left( {\frac{\gamma }{\mu }} \right)}^{\frac{1}{r}}}} \right)}^z}}}{{}}d} \gamma .
\end{aligned}
\end{equation}
In (\ref{BERuI1}), by letting $t = \ln \left( {\frac{1}{{{A_o}{{\left( {\frac{z}{{z + 1}}} \right)}^k}{{\left( {\frac{\gamma }{\mu }} \right)}^{\frac{1}{r}}}}}} \right)$, then $\gamma  = \mu {\left( {\frac{{{e^{ - t}}}}{{{A_o}{{\left( {\frac{z}{{z + 1}}} \right)}^k}}}} \right)^r}$ and $d\gamma  = \left( { - \mu r} \right){\left( {\frac{{{e^{ - t}}}}{{{A_o}{{\left( {\frac{z}{{z + 1}}} \right)}^k}}}} \right)^r}dt$. Thus, (\ref{BERuI1}) can be rewritten as follows:
\begin{equation} \label{BERuI1_1}
\begin{aligned}
{{\overline {BER} _{u,{\rm I}}} }&= \sum\limits_{l = 0}^\infty  {\frac{{{a_u}{{\left( { - \frac{{{b_u}}}{{{M_u}}}} \right)}^l}}}{{l!}}\cdot{\textstyle{1 \over {{F_\gamma }\left( {{\textstyle{\mu  \over {{{\left( {{A_o}{{\left( {\frac{z}{{z + 1}}} \right)}^k}} \right)}^r}}}}} \right) - {F_\gamma }\left( {{\gamma _T}} \right)}}}} \\
&\times \frac{{r{z^k}{\mu ^{l - 1}}}}{{\Gamma (k)}}{\left( {\frac{1}{{{{{{A_o}{{\left( {\frac{z}{{z + 1}}} \right)}^k}}}}}}} \right)^{r\left( {l - 1} \right)}}\int_{0}^{\ln \left( {{\textstyle{1 \over {{A_o}{{\left( {\frac{z}{{z + 1}}} \right)}^k}{{\left( {\frac{{{\gamma _{{T_u}}}}}{\mu }} \right)}^{\frac{1}{r}}}}}}} \right)} {{t^{k - 1}}{e^{ - t\left( {z - r + rl} \right)}}d} t.
\end{aligned}
\end{equation}
From (\ref{BERuI1_1}), utilizing \cite[Eq.(3.381.1)]{table}, the closed-form of the unified ${\overline {BER} _{u,{\rm I}}}$ in (\ref{BERuI1_1}) can be obtained as follows: 
\begin{equation} \label{BERIf}
\begin{aligned}
{\overline {BER} _{u,{\rm I}}} &= \sum\limits_{l = 0}^\infty  {\frac{{{a_u}{{\left( { - \frac{{{b_u}}}{{{M_u}}}} \right)}^l}}}{{l!}}{\textstyle{1 \over {{F_\gamma }\left( {{\textstyle{\mu  \over {{{\left( {{A_o}{{\left( {\frac{z}{{z + 1}}} \right)}^k}} \right)}^r}}}}} \right) - {F_\gamma }\left( {{\gamma _T}} \right)}}}} \\
&{\times \frac{{r{z^k}{\mu ^{l - 1}}}}{{\Gamma (k)}}{\left( {\frac{1}{{{{{{A_o}{{\left( {\frac{z}{{z + 1}}} \right)}^k}}}}}}} \right)^{r\left( {l - 1} \right)}}{\left( {z - r + rl} \right)^{ - k}}\gamma \left( {k,\left( {z - r + rl} \right)\ln \left( {{\textstyle{1 \over {{A_o}{{\left( {\frac{z}{{z + 1}}} \right)}^k}{{\left( {\frac{{{\gamma _{{T_u}}}}}{\mu }} \right)}^{\frac{1}{r}}}}}}} \right)} \right)}
\end{aligned}
\end{equation}
However, to obtain the closed form results of ${\overline {BER} _{u}}$ in both general cases and the case where pointing error is negligible, we still need to derive the closed-form results of ${\overline {BER} _{u,{\rm I}{\rm I}}}$ and ${\overline {BER} _{u,{\rm I}{\rm I}{\rm I}}}$ corresponding to the relevant cases. 
In the case of ${\overline {BER} _{u,{\rm I}{\rm I}}}$ and ${\overline {BER} _{u,{\rm I}{\rm I}{\rm I}}}$, we can perform the same derivation process as ${\overline {BER} _{u,{\rm I}}}$ in both cases.
Consequently, we can obtain the closed-form of unified ${\overline {BER} _{u,{\rm I}{\rm I}}}$ and ${\overline {BER} _{u,{\rm I}{\rm I}{\rm I}}}$ by replacing ${\gamma _{{T_u}}}$ in (\ref{unifiedBERuI}) and (\ref{BERIf}) with ${\gamma _{{T_{u+1}}}}$ and ${\gamma _{{_{{T_{{N_{\max }}}}}}}}$ for Case 1 and ${\gamma _{{T_{8}}}}$ for Case 2, respectively.
Subsequently, we can finally obtain the closed form results of ${\overline {BER} _{u}}$ by subtracting the closed form of ${\overline {BER} _{u,{\rm I}}}$ and ${\overline {BER} _{u,{\rm I}{\rm I}}}$ and adding the closed form of ${\overline {BER} _{u,{\rm I}{\rm I}{\rm I}}}$.
}

\begin{table}
\centering
\caption{Parameter settings in results}
\label{parameter}
\begin{tabular}{|c|c|c|c|}
\noalign{\smallskip}\noalign{\smallskip}\hline
Parameter & Symbol & Value \\
\hline
Distance & $L$ & 500m \\
\hline
Beamwidth at $L$ & $\omega _L$ & $0.3$m \\
\hline
Radius of receiver lens & ${r_o}$ & $0.1$m \\
\hline
SNR threshold& ${\gamma_T}$ & $14.0$ dB \\
\hline
\end{tabular}
\end{table}

{\section{Numerical results}}
In this section, we present performance outcomes to illustrate how applying TMOS-based WDM with ACM enhances the performance of FSO communication among mobile platforms in maritime settings amidst random fog and pointing inaccuracies. 
Our methodology involves selecting beams that surpass a predetermined SNR threshold (i.e., above $\gamma_T$). Based on this selection, we employed a rate-adaptive scheme utilizing an $N$-dimensional trellis coded M-QAM for our performance evaluation framework. 
More specifically, we define $N+1$ regions as the SNR threshold (i.e., $0 < {\gamma _{{T_1}}} < {\gamma _{{T_2}}} <  \cdots < {\gamma _{{T_N}}}<{\gamma _{{T_{N + 1}}}} $) and each region is assigned different modulation size $M_n$, $M_n={2^{n + 1}} \left( {n = 1,2, \ldots ,N} \right)$ for $N=8$ different codes based on QAM signal.
Note that ${\gamma }$ has an upper limit (i.e., ${\gamma _{{T_{N + 1}}}}={\textstyle{\mu  \over {{{\left( {{{\left( {\frac{z}{{z + 1}}} \right)}^k}\frac{\xi }{{\sqrt {\left( {1 + q\xi } \right)\left( {1 + \frac{\xi }{q}} \right)} }}} \right)}^r}}}}$), unlike  ${\gamma _{{T_{N + 1}}}}=\infty$ in the conventional \cite{Sung2017}.

{The means of random vector ${\bf{r}}$, ${{\boldsymbol{\mu }}_{\bf{r}}}$, mentioned in \ref{pointing error}} is expressed in spherical coordinates as $\left( {L,{\alpha _d},{\beta _d}} \right)$, i.e., ${\mu _x} = L\sin {\beta _d} \cos {\alpha _d} $, ${\mu _y} = L\sin {\beta _d} \sin {\alpha _d} $, and ${\mu _z} = L\cos {\beta _d} $ to quantify the effect of non-orthogonality of the beam, i.e., $\left( {{\alpha _d},{\beta _d}} \right) = \left( {\frac{\pi }{8},\frac{{5\pi }}{8}} \right)$ \cite{MarziehUAV}. 
Also, we assume standard deviations for the position and orientation variables as $\left( {{\sigma _x},{\sigma _y},{\sigma _z}} \right) = \sigma {r_0}\left( {0.8,0.27,0.53} \right)$ and $\left( {{\sigma _\theta },{\sigma _\phi }} \right) = \frac{{\sigma {r_0}}}{L}\left( {0.44,0.9} \right)$, respectively, where $\sigma $ is the standard deviation of the position or orientation \cite{MarziehUAV}.
Moreover, the analytical expressions presented in the previous sections are cross-verified using computer-based Monte-Carlo simulations under varying foggy condition parameters (i.e., $k$ and $\beta$) and pointing error parameters (i.e., $\xi$).
In all figures, the lines and markers represent the analytical and simulation results, respectively, and the analytical results perfectly match the simulation results.
The default parameter settings for all figures are listed in Table \ref{parameter} unless otherwise noted.
In all figures, we consider light fog with a low pointing error that slightly degrades the performance (i.e., $k=2.32$, $\beta=13.12$, and $\xi=420.7725$ with $\sigma=0.5$) and dense fog with a high pointing error that severely degrades the performance (i.e., $k=36.05$, $\beta=11.91$, and $\xi=0.6732$ with $\sigma=2.5$) and call them weak impact cases and severe impact cases, respectively.
Note that the higher $\xi$ value, the lower the pointing error effect.

\begin{figure}[ht] % float placement: (h)ere, page (t)op, page (b)ottom, other (p)age
  \centering
  % file name: C:/Users/lovej/Desktop/Task2fig/PDF.eps
  \includegraphics[width=10.0 cm]{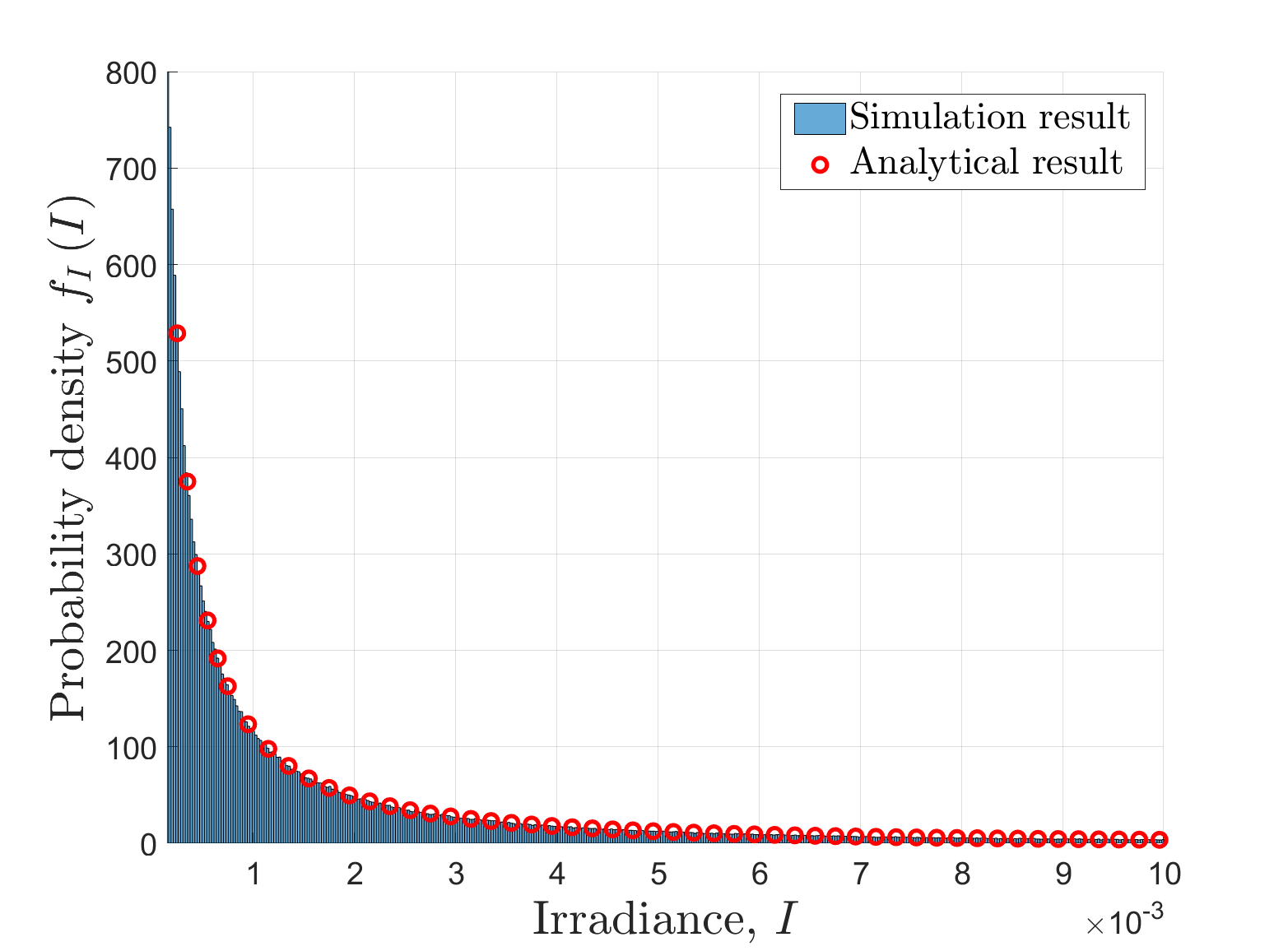}
  \caption{PDF comparisons for moderate fog with pointing error (i.e., $\sigma=1$).}
  \label{fig:PDF}
\end{figure}

In Fig. \ref{fig:PDF}, we present the composite PDF results, based on the system model detailed in Sec. \ref{systemmodel}. Specifically, this figure shows a comparative analysis between the simulation results and newly derived analytical outcomes, considering an environment characterized by moderate fog and pointing errors. In conclusion, this analysis confirmed that these two results matched perfectly.
\begin{figure}[ht] % float placement: (h)ere, page (t)op, page (b)ottom, other (p)age
  \centering
  % file name: C:/Users/lovej/Desktop/Task2fig/Outageprobabilitypreset.eps
  \includegraphics[width=10.0 cm]{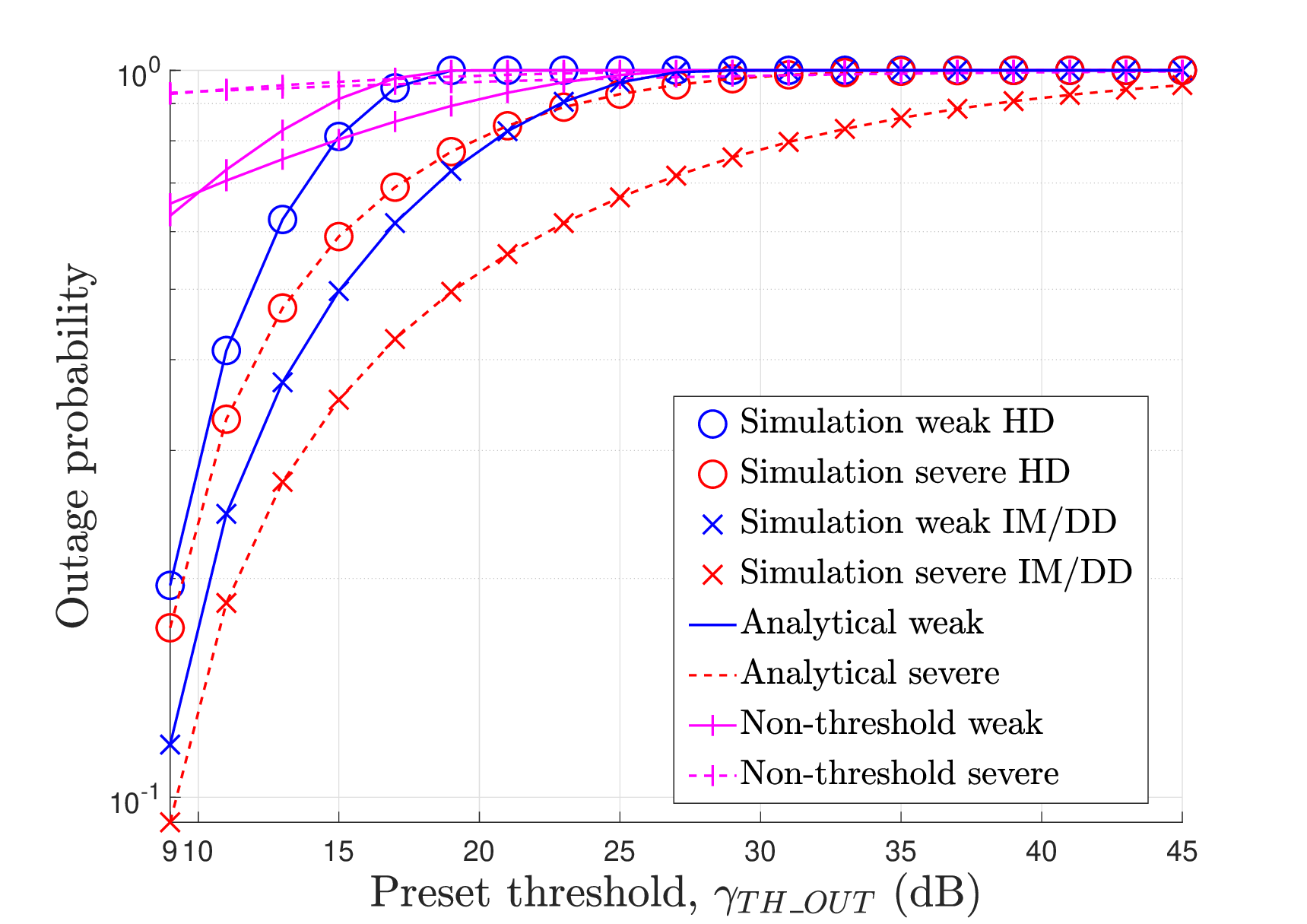}
  \caption{{Impact of an adjustable threshold on the outage probability in TMOS-based WDM and non-threshold-based WDM for combined effects of foggy channel with pointing error under both HD and IM/DD techniques with $\gamma_{T}=7.1$ dB.}}.
  \label{fig:Outageprobabilitypreset}
\end{figure}

\begin{figure}[ht] % float placement: (h)ere, page (t)op, page (b)ottom, other (p)age
  \centering
  % file name: C:/Users/lovej/Desktop/Task2fig/Outageprobabilitynon2.eps
  \includegraphics[width=10.0 cm]{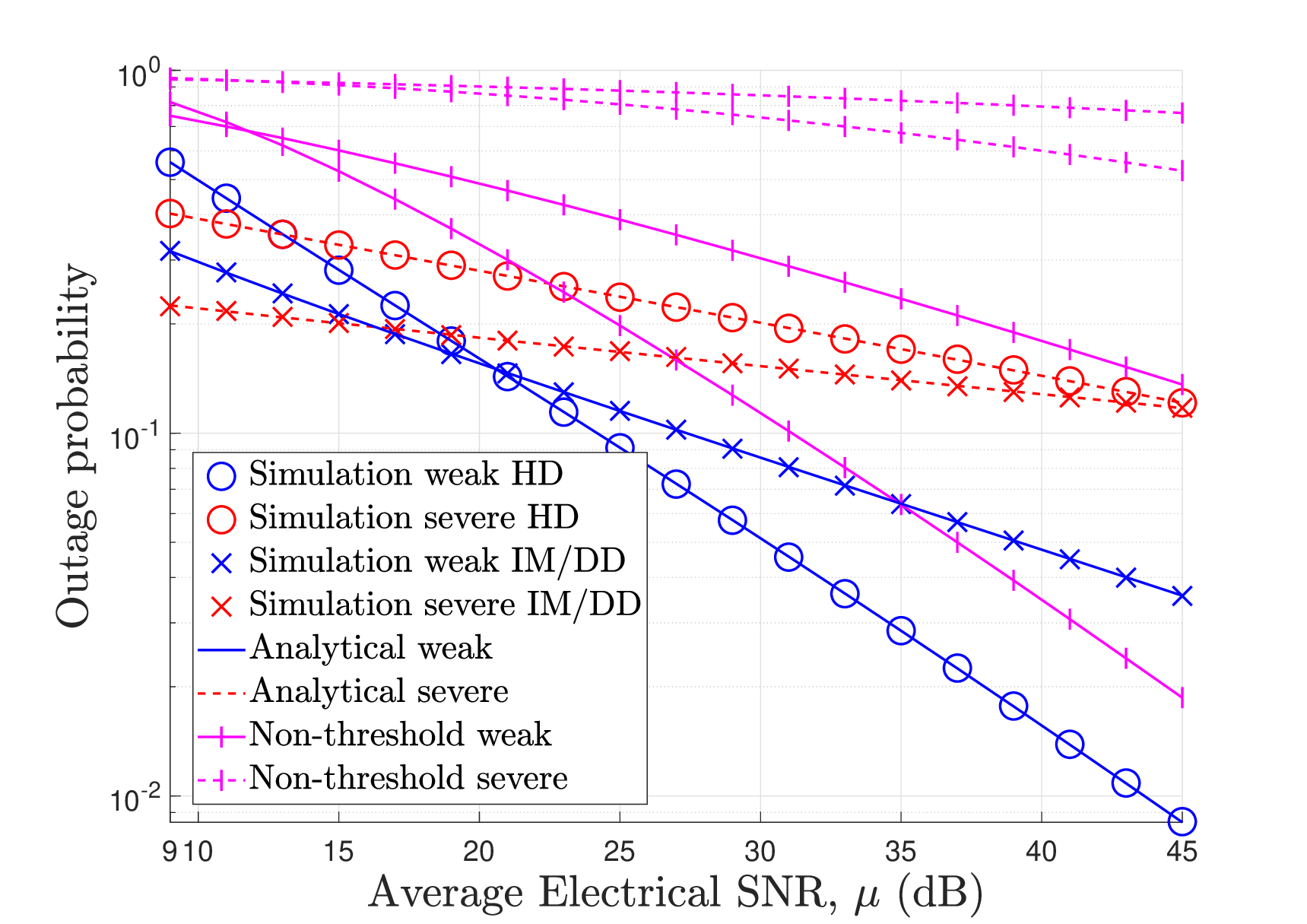}
  \caption{{Impact of varying $\mu$ on the outage probability in TMOS-based WDM and non-threshold-based WDM for combined effects of the foggy channel with pointing error under both HD and IM/DD techniques with $\gamma_{T}=7.1$ dB and $\gamma_{TH_\_{OUT}}=11.8$ dB.}}.
  \label{fig:Outageprobabilitynon2}
\end{figure}

{Fig. \ref{fig:Outageprobabilitypreset} and Fig. \ref{fig:Outageprobabilitynon2} display the outage probabilities of FSO systems using i) TMOS-based WDM and ii) non-threshold-based WDM, under both HD and IM/DD techniques with $\gamma_{T}=7.1$ dB . 
These figures illustrate the performances of these systems under varying levels of influencing factors, from weak to severe conditions. 
Specifically, Fig. \ref{fig:Outageprobabilitypreset} shows the impact of an adjustable threshold on the outage probability, with $\mu$ (i.e., $\mu=10$ dB) being held constant, as described in Eq. (\ref{POUT}). 
In this context, an $"$outage$"$ refers to situations where the quality of the link for each selected beam falls below the threshold specified in Eq. (\ref{POUT}). 
As is evident from the outcomes, it is expected that when $\mu$ is fixed at $10$ dB, increasing the threshold leads to a reduction in the number of beams with SNR above the threshold. 
This reduction, in turn, results in the deterioration of FSO communication.
Fig. \ref{fig:Outageprobabilitynon2} shows the impact of varying $\mu$ on the probability of successful data transmission at a fixed $\gamma_{TH_\_{OUT}}=11.8$ dB. 
In this scenario, with increasing values of $\mu$, there is a corresponding increase in the number of beams with SNR higher than the threshold, resulting in improved system performance. 
Furthermore, the results confirm that the performance of TMOS-based WDM surpasses that of non-threshold-based WDM, particularly in challenging maritime environments. 
Essentially, by employing the threshold, beams with SNR compared to the threshold that fail to meet the system requirements are filtered out, leaving only those beams that satisfy the system requirements, thus enhancing beam reliability.

\begin{figure}[ht] % float placement: (h)ere, page (t)op, page (b)ottom, other (p)age
  \centering
  % file name: C:/Users/lovej/Desktop/Task2fig/ANSB.eps
  \includegraphics[width=10.0 cm]{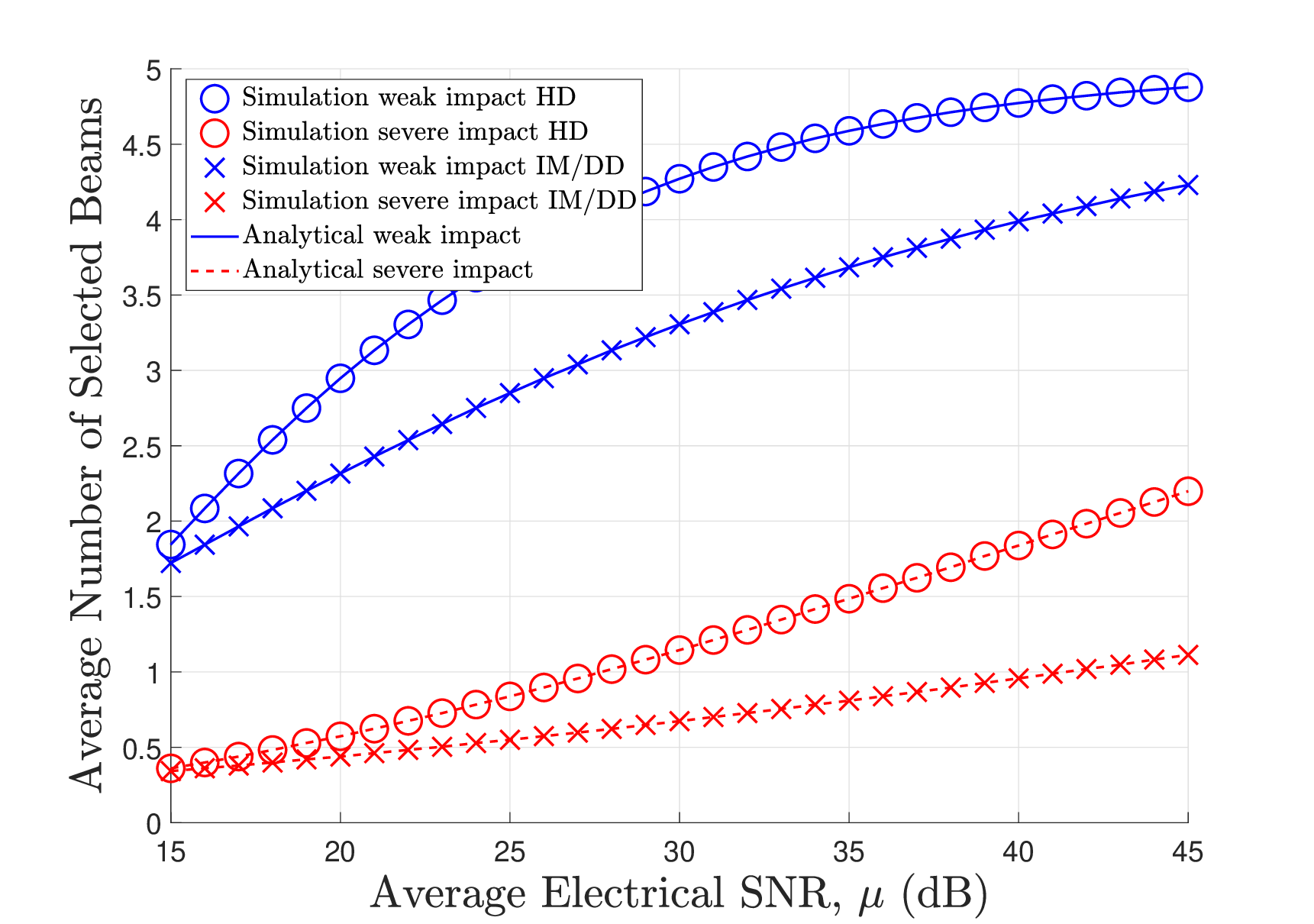}
  \caption{Average number of selected beams for combined effects of foggy channel with pointing error under both HD and IM/DD techniques with $\gamma_{T}=14.0$ dB.}
  \label{fig:ANSBth1420230205}
\end{figure}

\begin{figure}[ht] % float placement: (h)ere, page (t)op, page (b)ottom, other (p)age
  \centering
  % file name: C:/Users/lovej/Desktop/Task2fig/ASEbeams.eps
  \includegraphics[width=10.0 cm]{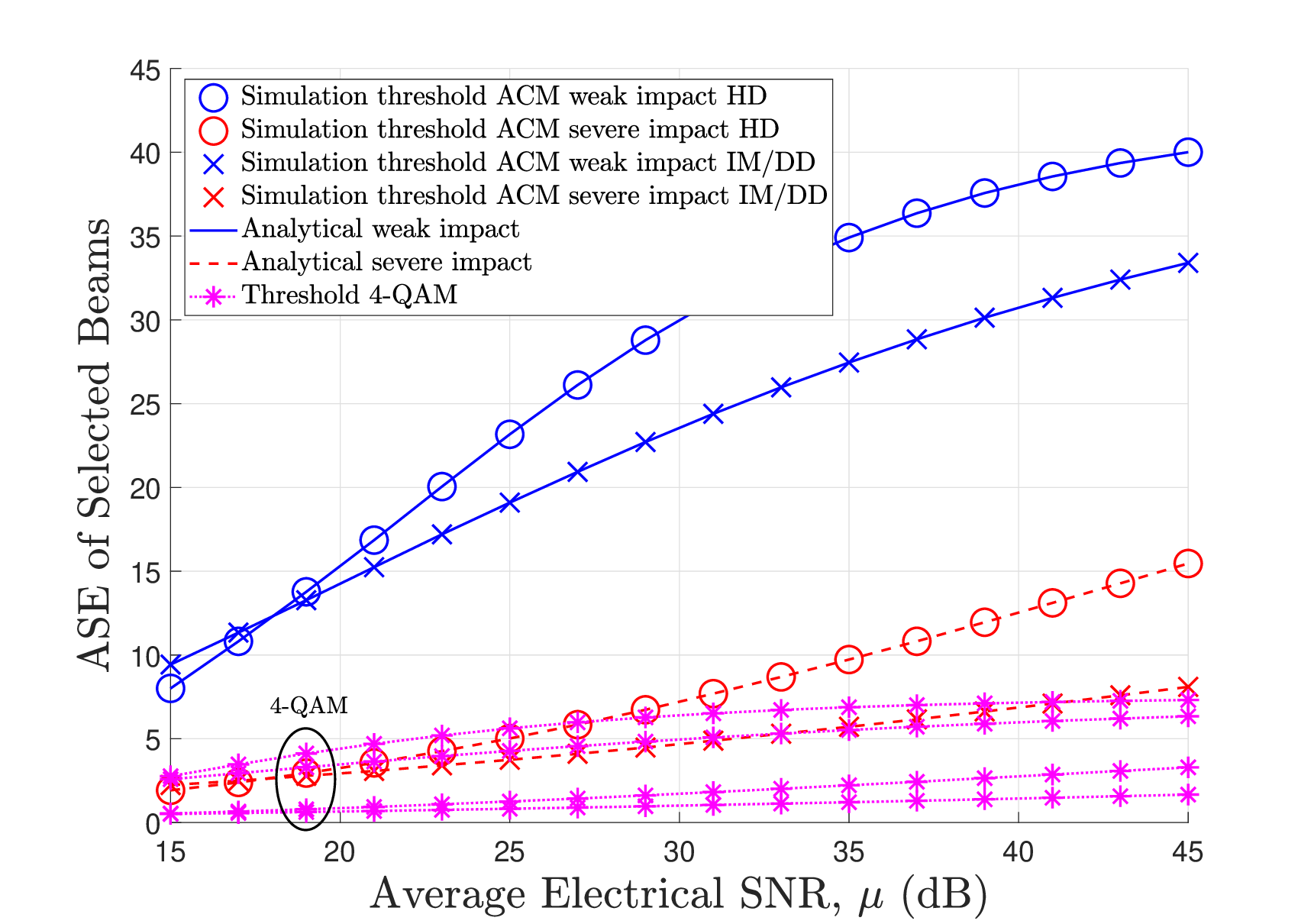}
  \caption{Average spectral efficiency of selected beams in TMOS-based WDM, both with and without ACM, for combined effects of foggy channel with pointing error under both HD and IM/DD techniques with $\gamma_{T}=14.0$ dB.}
  \label{fig:ASEth144QAM20230205}
\end{figure}

\begin{figure}[ht] % float placement: (h)ere, page (t)op, page (b)ottom, other (p)age
  \centering
  % file name: C:/Users/lovej/Desktop/Task2fig/AverageBER.eps
  \includegraphics[width=10.0 cm]{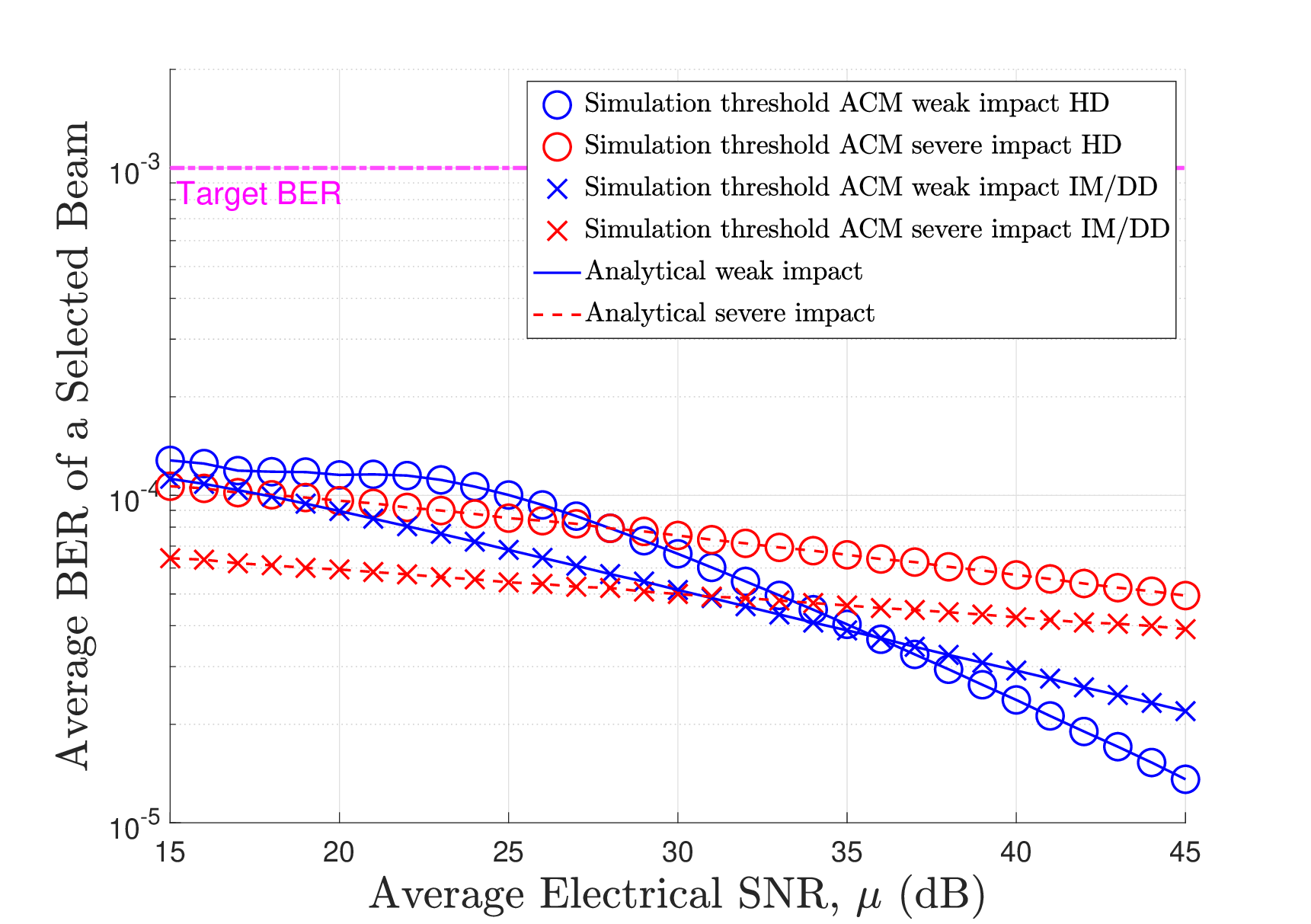}
  \caption{Average BER of a selected beam in TMOS-based WDM with ACM for combined effects of foggy channel with pointing error under both HD and IM/DD techniques with $\gamma_{T}=14.0$ dB.}
  \label{fig:AverageBERth1420230205}
\end{figure}

%%%%%%%%%%%%%%%%%%%%%%%%%%%%%%%%%%%%%%
%%%%%%%%%%%%%%%%%%%%%%%%%%%%%%%%%%%%%%

\begin{figure}[ht] % float placement: (h)ere, page (t)op, page (b)ottom, other (p)age
  \centering
  % file name: C:/Users/lovej/Desktop/Task2fig/AverageBERmix.eps
  \includegraphics[width=10.0 cm]{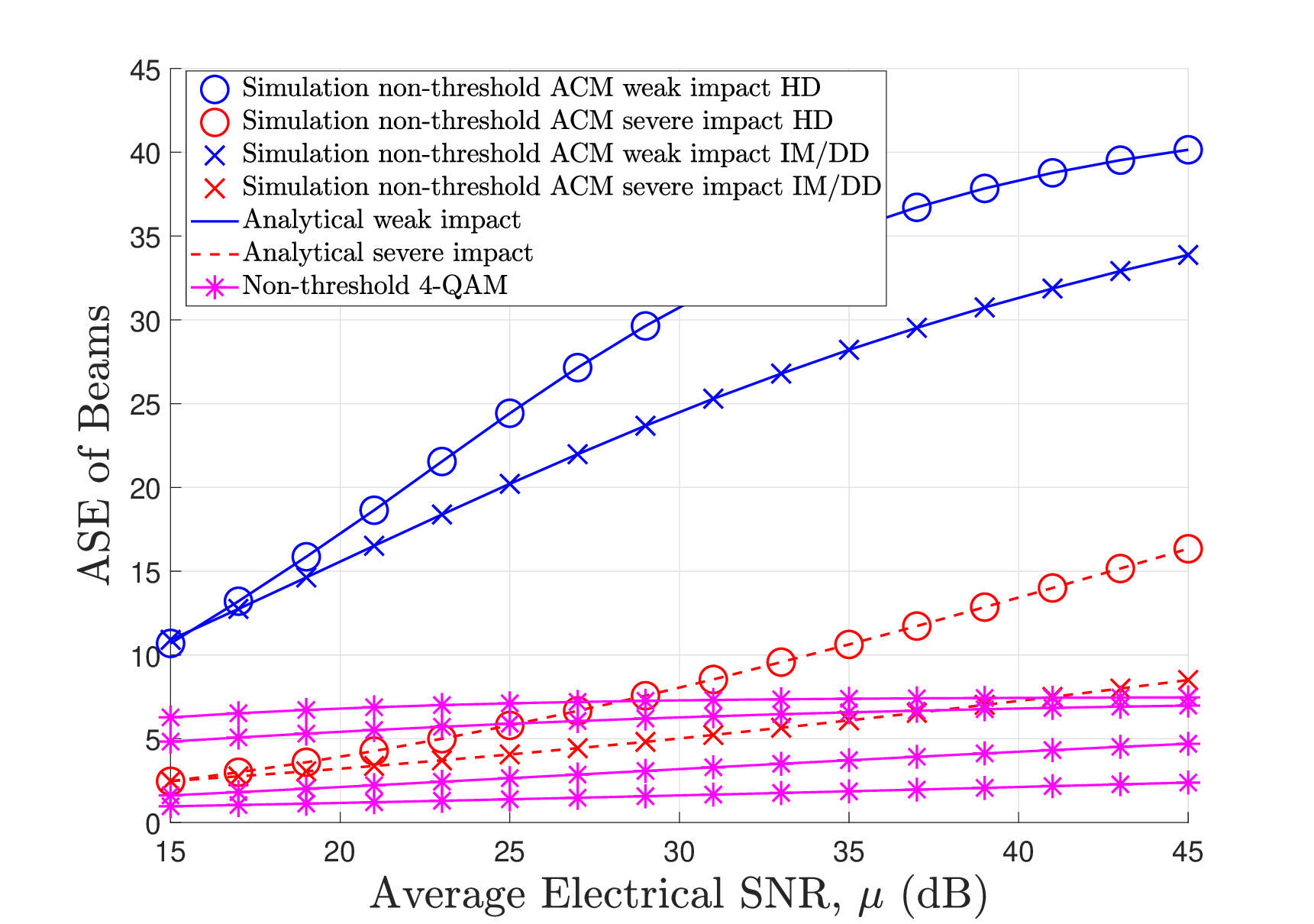}
  \caption{Average spectral efficiency of beams in non-threshold-based WDM, both with and without ACM, for combined effects of foggy channel with pointing error under both HD and IM/DD techniques.}
  \label{fig:ASEbeamsnonth20230301}
\end{figure}

\begin{figure}[ht] % float placement: (h)ere, page (t)op, page (b)ottom, other (p)age
  \centering
  % file name: C:/Users/lovej/Desktop/Task2fig/AverageBERmix.eps
  \includegraphics[width=10.0 cm]{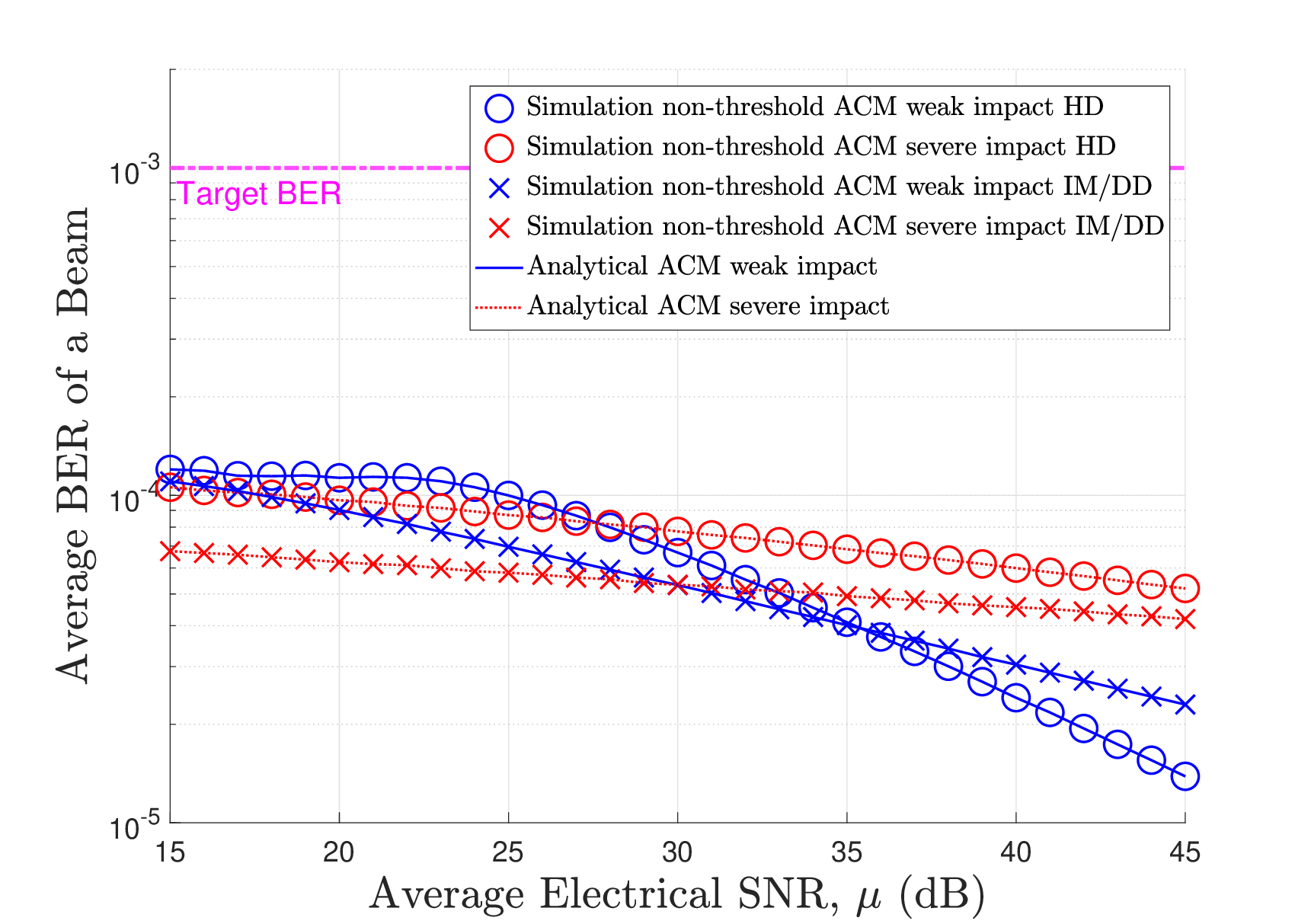}
  \caption{Average BER of a beam in non-threshold-based WDM with ACM for combined effects of foggy channel with pointing error under both HD and IM/DD techniques.}
  \label{fig:AverageBERnonth20230205}
\end{figure}
}

Fig. \ref{fig:ANSBth1420230205} shows the average number of selected beams (ANSB) with TMOS-based WDM in both the weak and severe impact case under HD and IM/DD techniques  with $\gamma_{T}=14.0$ dB.
Expectedly, as $\mu$ increases, the number of beams above the predefined threshold increases; thus, ANSB is enhanced and eventually converges to the total beam number (i.e., $H=5$). In addition, for both detection techniques, the weak impact case performed better than the severe impact case and converged much faster. 
Based on these results, Fig. 8 depicts the ASE of the selected beams in TMOS-based WDM, both with and without ACM, in scenarios with weak or severe impacts on the communication channel with $\gamma_{T}=14.0$ dB. We assume that, in the context of fixed modulation, which consistently transmits data at the same rate, it is typically designed to operate at the minimum SNR, typically using the $4$-QAM scheme (i.e., $R_u=1.5$ in $4$-QAM). When adaptive modulation is adopted, the data rate is dynamically adjusted based on the channel conditions, allowing for the transmission of more data compared with fixed modulation. Consequently, as the SNR increases, the spectral efficiency also increases, leading to an upward trend in ASE, ultimately approaching the maximum achievable ASE $L \cdot {R_N}$ (i.e., $5 \times 8.5 = 42.5$). In systems employing ACM, the primary goal is to optimize the data transmission rates while maintaining a stable BER as required by the system in response to the current channel conditions. The results presented in Figure 9 confirm that the average BER across a range of scenarios was consistently maintained below the predefined threshold BER. ($\overline {BE{R_0}} = {10^{ - 3}}$), this criterion is maintained even in the face of significant channel degradation. Consequently, the utilization of ACM in dynamically changing channels not only meets the desired system quality but also boosts the ASE. This demonstrates that data transmission is considerably more efficient than when using fixed modulation schemes.

To assess the impact of applying the threshold-based WDM, Fig. \ref{fig:ASEbeamsnonth20230301} and Fig. \ref{fig:AverageBERnonth20230205} depict the ASE and average BER when using conventional (i.e., non-threshold-based) WDM with ACM in scenarios of both weak and severe signal degradation, considering both HD and IM/DD techniques. Without considering the threshold, the number of beams remains constant for the total number of beams (i.e., $5$), which leads to variations in the performance of ANSB, as illustrated in Fig. \ref{fig:ASEbeamsnonth20230301}. However, the disparity between the ASE shown in Fig. \ref{fig:ASEth144QAM20230205} and that in Fig. \ref{fig:ASEbeamsnonth20230301} is very small. This indicates that beams with SNRs falling below the threshold are ill-suited for communication and have an insignificant impact on system performance. Furthermore, Fig. 11 confirms that in scenarios where the threshold is not considered, all outcomes result in BER lower than the specified target. Consequently, by considering the threshold-based selection scheme with ACM, with fewer beams used for transmission, a performance similar to that achieved when threshold-based WDM is not considered is still attainable.

Figures \ref{fig:Outageprobabilitynon2}, \ref{fig:AverageBERth1420230205}, and \ref{fig:AverageBERnonth20230205}  demonstrate the improved performance in scenarios where the communication channel is severely impacted at a low SNR, which differs from typical outcomes. This can be attributed to the combined effect of the presence of fog and jitter. The beam experiences dispersion as it traverses the foggy channel, as discussed in \cite{Anbarasi}. Furthermore, the jitter makes the beam footprint more spread out \cite{MarziehUAV}. Owing to these factors, a dispersed beam has an elevated likelihood of being detected using a photodetector (PD). Notably, in the scenario with a severe impact on the communication channel, where the extent of beam widening is more pronounced than in the weak impact case, the probability of detection increases, resulting in enhanced system performance.

However, as the SNR increases, the beam spread relatively diminishes, resulting in a reduction in the positive impact on the performance caused by the combined effects of beam spread and jitter. Generally, FSO operates across a broader range of SNR than RF, and notably functions at high SNRs \cite{Sung2017}. In this context, when focusing on a higher SNR range, the HD technique more effectively mitigates the adverse impact of fog and pointing errors compared to the IM/DD technique, leading to improved performance. This disparity in effectiveness stems from the fundamental operational differences between the two techniques: HD technique implements a two-dimensional modulation involving phase and amplitude, whereas the IM/DD technique operates within one dimension, focusing solely on intensity \cite{Anas}. Therefore, IM/DD technique necessitates a higher SNR than its HD counterpart to achieve equivalent performance rates.

Note that when $\mu$ is $45$ dB, the average SNR affected by the weak and severe impact cases are $39.4631$ dB in HD and $38.0607$ dB in IM/DD, and $27.7026$ dB in HD and $32.4938$ dB in IM/DD, respectively. 
Here, it is possible to apply the conventional values of the coefficients (i.e., $a_u$ and $b_u$) used in \cite{Kjell}.
Even if ACM is applied to a wider and higher range of SNR, the analytical framework and closed-form expression results, along with the appropriate coefficient values provided in \cite{Kjell}, remain valid.
However, suitable values of $a_n$ and $b_n$ should be determined using curve-fitting techniques based on  $M_u$, ${\gamma}_{T_u}$, $R_u$, and $\overline {BE{R_0}} $ using the least squares method, as shown in \cite{Kjell}.
Furthermore, according to \cite{Sung2017}, considering the pathloss ${I_l}$ to be deterministic, we assume that it maintains a constant value of $1$.
However, if the impact of the pathloss needs to be considered, we can substitute ${A_0}$ with ${A_0} \cdot {I_l} $ in the derived results.

\section{Conclusion}
In this study, we considered FSO communication between mobile platforms in maritime environments where fog and 3D pointing errors are the major performance degradations and applied TMOS-based WDM and ACM to improve the FSO performance.
We derived a new closed-form expression for a composite PDF that is particularly suitable for statistical performance analysis when various algorithms are applied to maritime FSO systems under the combined effects of fog and pointing errors in maritime environments. This new result is particularly suitable for statistical performance analyses when various algorithms are applied to maritime FSO systems under the combined effects of fog and pointing errors in maritime environments. This contribution is pivotal in advancing the domain by improving statistical analysis and performance evaluation capabilities in the context of maritime communications under challenging environmental conditions.
We then analyzed the outage probability, ASE, and BER performance analysis for both conventional detection techniques (i.e., HD and IM/DD).
All analytical results are comparatively validated and fully matched through Monte Carlo simulations.
Based on the selected results, we can confirm that a higher ASE performance can be achieved by adjusting the data rate according to the channel environment by applying TMOS-based WDM and ACM.
Therefore, it was confirmed that the application of TMOS-based WDM and ACM is suitable for maritime environments with frequent changes in channel conditions.
In addition, the performance difference between HD and IM/DD techniques can be identified, which can help determine an appropriate detection technique. 
However, in some results, the performance was reversed at a low SNR unlike the conventional performance, which confirmed that the performance improved by increasing the probability that the diffused beam was detected in the PD owing to the combined effects of fog and jitter.
Note that this contribution is pivotal in advancing the domain by improving statistical analysis and performance evaluation capabilities in the context of maritime communications under challenging environmental conditions.

{
\appendices
\section{Composite PDF}
\label{composite PDF}
By substituting (\ref{iapdf}) and (\ref{ippdf}) into (\ref{pdfI}), the composite PDF can be written as follows: 
\begin{equation} \label{pdfIsub}
\begin{aligned}
{f_I}\left( I \right) =& \int_{\frac{I}{{{A_0}}}}^1 {\frac{{{z^k}}}{{\Gamma \left( k \right)}}\frac{\xi }{{{A_0}}}{{\left( {\frac{I}{{{A_0}}}} \right)}^{\frac{{\left( {1 + {q^2}} \right)\xi }}{{2q}} - 1}}{I_a}^{ - \frac{{\left( {1 + {q^2}} \right)\xi }}{{2q}} + z - 1}   }\\
&\times{  {I_0}\left( { - \frac{{\left( {1 - {q^2}} \right)\xi }}{{2q}}\ln \left( {\frac{I}{{{I_a}{A_0}}}} \right)} \right){{\left[ {\ln \left( {\frac{1}{{{I_a}}}} \right)} \right]}^{k - 1}}d{I_a}} .
\end{aligned}
\end{equation}
Then, using \cite[Eq. (8.447.1)]{table}, (\ref{pdfIsub}) can be expressed as follows: 
\begin{equation} \label{pdfIsubsub}
\begin{aligned}
{f_I}\left( I \right) =& \int_{\frac{I}{{{A_0}}}}^1 {\frac{{{z^k}}}{{\Gamma \left( k \right)}}\frac{\xi }{{{A_0}}}{{\left( {\frac{I}{{{A_0}}}} \right)}^{\frac{{\left( {1 + {q^2}} \right)\xi }}{{2q}} - 1}}{I_a}^{ - \frac{{\left( {1 + {q^2}} \right)\xi }}{{2q}} + z - 1}    }\\
&\times{   \sum\limits_{m = 0}^\infty  {\frac{1}{{{{\left( {m!} \right)}^2}}}} {{\left( { - \frac{{\left( {1 - {q^2}} \right)\xi }}{{2q}}\frac{1}{2}\ln \left( {\frac{I}{{{I_a}{A_0}}}} \right)} \right)}^{2m}}{{\left[ {\ln \left( {\frac{1}{{{I_a}}}} \right)} \right]}^{k - 1}}d{I_a}} .
\end{aligned}
\end{equation}
Here, let $x = \frac{1}{2}\ln \left( {\frac{{{I_a}{A_0}}}{I}} \right)$, then ${I_a} = \frac{I}{{{A_0}}}\exp \left( {2x} \right)$ and $d{I_a} = \frac{{2I}}{{{A_0}}}\exp \left( {2x} \right)dx$, and the integral region of $x$ becomes $0 < x \le \frac{1}{2}\ln \left( {\frac{{{A_0}}}{I}} \right)$.
Then, (\ref{pdfIsubsub}) can be expressed as follows:
\begin{equation}
\begin{aligned} \label{pdfIsubsubsub}
{f_I}\left( x \right) &= \frac{{2{z^k}}}{{\Gamma \left( k \right)}}\frac{\xi }{{{A_0}}}{\left( {\frac{I}{{{A_0}}}} \right)^{z - 1}}\sum\limits_{m = 0}^\infty  {\frac{1}{{{{\left( {m!} \right)}^2}}}{{\left( { - \frac{{\left( {1 - {q^2}} \right)\xi }}{{2q}}} \right)}^{2m}}   }\\
&\times{  \int_0^{\frac{1}{2}\ln \left( {\frac{{{A_0}}}{I}} \right)} {{x^{2m}}{\exp \left( x\left( { - \frac{{\left( {1 + {q^2}} \right) \xi }}{{q}} + 2z} \right)\right)}} {{\left[ {\ln \left( {\frac{{{A_0}}}{I}} \right) - 2x} \right]}^{k - 1}}dx} .
\end{aligned}
\end{equation}
In (\ref{pdfIsubsubsub}), let $\varpi  =  - \frac{{\left( {1 + {q^2}} \right)\xi }}{{q}} + 2z$ and $B = \frac{1}{2}\ln \left( {\frac{{{A_0}}}{I}} \right)$, then (\ref{pdfIsubsubsub}) can be written as follows:
\begin{equation}
\begin{aligned}\label{pdfIsubsubsubsub}
{f_I}\left( x \right) &= \frac{{2{z^k}}}{{\Gamma \left( k \right)}}\frac{\xi }{{{A_0}}}{\left( {\frac{I}{{{A_0}}}} \right)^{z - 1}}\sum\limits_{m = 0}^\infty  {\frac{1}{{{{\left( {m!} \right)}^2}}}{{\left( { - \frac{{\left( {1 - {q^2}} \right)\xi }}{{2q}}} \right)}^{2m}}    }\\
&\times{   \int_0^B {{x^{2m}}\exp \left( {\varpi x} \right){{\left( {2B} \right)}^{k - 1}}{{\left( {1 - \frac{x}{B}} \right)}^{k - 1}}dx} } .
\end{aligned}
\end{equation}
In (\ref{pdfIsubsubsubsub}), using a binomial expansion \cite[Eq. (1.110)]{table}, the (\ref{pdfIsubsubsubsub}) can be expressed as follows: 
\begin{equation}
\begin{aligned} \label{pdfIsubsubsubsubsub}
{f_I}\left( x \right) &= \frac{{2{z^k}}}{{\Gamma \left( k \right)}}\frac{\xi }{{{A_0}}}{\left( {\frac{I}{{{A_0}}}} \right)^{z - 1}}\sum\limits_{m = 0}^\infty  {\frac{1}{{{{\left( {m!} \right)}^2}}}{{\left( { - \frac{{\left( {1 - {q^2}} \right)\xi }}{{2q}}} \right)}^{2m}}{{\left( {2B} \right)}^{k - 1}}    }\\
&\times{ \sum\limits_{n = 0}^\infty  \binom{k-1}{n} {{\left( { - \frac{1}{B}} \right)}^n}\int_0^B {{x^{2m + n}}\exp \left( {\varpi x} \right)dx} } .
\end{aligned}
\end{equation}
In (\ref{pdfIsubsubsubsubsub}), by using \cite[Eq. (3.351.1)]{table}, the closed-form expression of (\ref{pdfIsubsubsubsubsub}) can be written as follows: 
\begin{equation}
\begin{aligned} \label{pdfIsubsubsubsubsubsub}
{f_I}\left( I \right) &= \frac{{2{z^k}}}{{\Gamma \left( k \right)}}\frac{\xi }{{{A_0}}}{\left( {\frac{I}{{{A_0}}}} \right)^{z - 1}}\sum\limits_{m = 0}^\infty  {\frac{1}{{{{\left( {m!} \right)}^2}}}{{\left( { - \frac{{\left( {1 - {q^2}} \right)\xi }}{{2q}}} \right)}^{2m}}{{\left( {2B} \right)}^{k - 1}} }\\
&\times{  \sum\limits_{n = 0}^\infty \binom{k-1}{n} {{\left( { - \frac{1}{B}} \right)}^n}{{\left( { - \varpi } \right)}^{ - (2m + n) - 1}}\gamma \left( {2m + n + 1, - \varpi B} \right)},
\end{aligned}
\end{equation}
where $\varpi  =  - \frac{{\left( {1 + {q^2}} \right)\xi }}{{q}} + 2z$ and $B = \frac{1}{2}\ln \left( {\frac{{{A_0}}}{I}} \right)$.

\section{Outage Probability based on HD technique}
\label{AppenoutageprobabilityHD} 
From (\ref{pdfsnrHD}), the outage probability of $\gamma$ for HD technique case can be written as follows:
\begin{equation}
\begin{aligned}\label{outage probabilitysnrHD}
 {F_\gamma }\left( x \right) &= \int_0^x {{f_\gamma }\left( \gamma  \right)d\gamma } \\
& = \frac{{2{z^k}}}{{\Gamma \left( k \right)}}\xi {\left( {\frac{{{{\left( {\frac{z}{{z + 1}}} \right)}^k}\frac{\xi }{{\sqrt {\left( {1 + q\xi } \right)\left( {1 + \frac{\xi }{q}} \right)} }}}}{{{\mu _{HD}}}}} \right)^z}\sum\limits_{m = 0}^\infty  {\frac{1}{{{{\left( {m!} \right)}^2}}}{{\left( { - \frac{{\left( {1 - {q^2}} \right)\xi }}{{2q}}} \right)}^{2m}}\sum\limits_{n = 0}^\infty  \binom{k-1}{n} } \\   
& {\hspace{12pt}} \times {\left( { - \varpi } \right)^{ - \left( {2m + n} \right) - 1}}{\left( { - 2} \right)^n}\int_0^x  {\gamma ^{z - 1}} {{\bigg( {\ln \bigg( {\frac{{{\mu _{HD}}}}{{\gamma {{\left( {\frac{z}{{z + 1}}} \right)}^k}\frac{\xi }{{\sqrt {\left( {1 + q\xi } \right)\left( {1 + \frac{\xi }{q}} \right)} }}}}} \bigg)} \bigg)}^{k - 1 - n}} \\
&{\hspace{12pt}}  \times  \gamma \bigg( 2m + n + 1, -\frac{\varpi }{2}
\ln \bigg( 
\frac{\mu _{HD}}{\gamma \left(\frac{z}{z+1} \right)^k \frac{\xi}
{\sqrt{\left( {1+q\xi} \right) 
\left({1 + \frac{\xi }{q}} \right)
}
}}\bigg) \bigg) 
d\gamma .
\end{aligned}
\end{equation}
Here, let $y = \ln \left( {\frac{{{\mu _{HD}}}}{{\gamma {{\left( {\frac{z}{{z + 1}}} \right)}^k}\frac{\xi }{{\sqrt {\left( {1 + q\xi } \right)\left( {1 + \frac{\xi }{q}} \right)} }}}}} \right)$, then $\gamma  = \frac{{{\mu _{HD}}}}{{{{\left( {\frac{z}{{z + 1}}} \right)}^k}\frac{\xi }{{\sqrt {\left( {1 + q\xi } \right)\left( {1 + \frac{\xi }{q}} \right)} }}}}\exp \left( { - y} \right)$, $d\gamma  = \frac{{ - {\mu _{HD}}}}{{{{\left( {\frac{z}{{z + 1}}} \right)}^k}\frac{\xi }{{\sqrt {\left( {1 + q\xi } \right)\left( {1 + \frac{\xi }{q}} \right)} }}}}\\
\cdot \exp \left( { - y} \right)dy$ , and the integral region of $y$ becomes $\ln \left( {\frac{{{\mu _{HD}}}}{{x {{\left( {\frac{z}{{z + 1}}} \right)}^k}\frac{\xi }{{\sqrt {\left( {1 + q\xi } \right)\left( {1 + \frac{\xi }{q}} \right)} }}}}} \right) < y < \infty $.
Then, (\ref{outage probabilitysnrHD}) can be expressed as follows:
\begin{equation}
\begin{aligned}\label{outage probabilitysnrHDHD}
 {F_\gamma }\left( y \right) 
& = \frac{{2{z^k}}}{{\Gamma \left( k \right)}}\xi \sum\limits_{m = 0}^\infty  {\frac{1}{{{{\left( {m!} \right)}^2}}}{{\left( { - \frac{{\left( {1 - {q^2}} \right)\xi }}{{2q}}} \right)}^{2m}}\sum\limits_{n = 0}^\infty  \binom{k-1}{n} } 
 {\left( { - \varpi } \right)^{ - \left( {2m + n} \right) - 1}}{\left( { - 2} \right)^n}\\
& \times\int_{\ln \left( {\frac{{{\mu _{HD}}}}{{x{{\left( {\frac{z}{{z + 1}}} \right)}^k}\frac{\xi }{{\sqrt {\left( {1 + q\xi } \right)\left( {1 + \frac{\xi }{q}} \right)} }}}}} \right)}^\infty   {\gamma \left( {2m + n + 1, - \frac{\varpi }{2}y} \right)\exp \left( { - zy} \right){y^{k - 1 - n}}dy}.
\end{aligned}
\end{equation}
The integral part in (\ref{outage probabilitysnrHDHD}) can be divided into two parts (i.e. $0 < y < \infty$ and $0 < y < \ln \left( {\frac{{{\mu _{HD}}}}{{x {{\left( {\frac{z}{{z + 1}}} \right)}^k}\frac{\xi }{{\sqrt {\left( {1 + q\xi } \right)\left( {1 + \frac{\xi }{q}} \right)} }}}}} \right)$.
Subsequently, using \cite[Eq. (6.455.2)]{table}, the first integral part (i.e., $0 < y < \infty$) can be expressed as follows:
\begin{equation}
\begin{aligned}
\int_0^\infty & {\gamma \left( {2m + n + 1, - \frac{\varpi }{2}y} \right)\exp \left( { - zy} \right){y^{k ? 1 ? n}}dy} \\
&= \frac{{{{\left( { - \frac{\varpi }{2}} \right)}^{2m + n + 1}}\Gamma \left( {k + 2m + 1} \right)}}{{\left( {2m + n + 1} \right){{\left( { - \frac{\varpi }{2} + z} \right)}^{k + 2m + 1}}}}\times{}_2{F_1}\left( {1,k + 2m + 1;2m + n + 2;\frac{{ - \varpi }}{{ - \varpi  + 2z}}} \right),
\end{aligned}
\end{equation} 
where,
with the help of theorem derived in Appendix \ref{Appendclosed}, the second integral part (i.e. $0 < y < \ln \left( {\frac{{{\mu _{HD}}}}{{x {{\left( {\frac{z}{{z + 1}}} \right)}^k}\frac{\xi }{{\sqrt {\left( {1 + q\xi } \right)\left( {1 + \frac{\xi }{q}} \right)} }}}}} \right)$) can be expressed as follows:
\begin{equation}
\begin{aligned}
   &\int_0^{\ln \left( {\frac{{{\mu _{HD}}}}{{x{{\left( {\frac{z}{{z + 1}}} \right)}^k}\frac{\xi }{{\sqrt {\left( {1 + q\xi } \right)\left( {1 + \frac{\xi }{q}} \right)} }}}}} \right)}{\gamma \left( {2m + n + 1, - \frac{\varpi }{2}y} \right)\exp \left( { - zy} \right){y^{k - 1 - n}}dy} \\
&= \sum\limits_{i = 0}^{k - 1 - n} {\frac{\left( {k - 1 - n} \right)!}{i!{z^{k - n -i }}}}\Bigg( 
  {{\left( { - \frac{\varpi }{2}} \right)}^{2m + n + 1}} {{\left( { - \frac{\varpi }{2} + z} \right)}^{ - (2m + n + 1) - i}}  \\
  & \times  \Bigg(
   {\gamma \bigg( {2m + n + 1 + i,\left( { - \frac{\varpi }{2} + z} \right)\ln \bigg( {\frac{{{\mu _{HD}}}}{{x{{\left( {\frac{z}{{z + 1}}} \right)}^k}\frac{\xi }{{\sqrt {\left( {1 + q\xi } \right)\left( {1 + \frac{\xi }{q}} \right)} }}}}} \bigg)} \bigg)}  \\
  &{ - \gamma \bigg( {2m + n + 1 , - \frac{\varpi }{2}\ln \bigg( {\frac{{{\mu _{HD}}}}{{x{{\left( {\frac{z}{{z + 1}}} \right)}^k}\frac{\xi }{{\sqrt {\left( {1 + q\xi } \right)\left( {1 + \frac{\xi }{q}} \right)} }}}}} \bigg)} \bigg){{\bigg( {\frac{{{\mu _{HD}}}}{{x{{\left( {\frac{z}{{z + 1}}} \right)}^k}\frac{\xi }{{\sqrt {\left( {1 + q\xi } \right)\left( {1 + \frac{\xi }{q}} \right)} }}}}} \bigg)}^{ - z}}} \\
  & \times {{{\bigg( {\ln \bigg( {\frac{{{\mu _{HD}}}}{{x{{\left( {\frac{z}{{z + 1}}} \right)}^k}\frac{\xi }{{\sqrt {\left( {1 + q\xi } \right)\left( {1 + \frac{\xi }{q}} \right)} }}}}} \bigg)} \bigg)}^i}}
 \Bigg)\Bigg).
\end{aligned}
\end{equation}

Consequently, the closed-form result of the outage probability of $\gamma$ for HD technique case can be obtained as follows:
\begin{equation}
\begin{aligned}
   {F_\gamma }\left( x  \right) &= \frac{{2{z^k}}}{{\Gamma \left( k \right)}}\xi {\sum\limits_{m = 0}^\infty  {\frac{1}{{{{\left( {m!} \right)}^2}}}\left( { - \frac{{\left( {1 - {q^2}} \right)\xi }}{{2q}}} \right)} ^{2m}}\sum\limits_{n = 0}^\infty \binom{k-1}{n} {{{\left( { - \varpi } \right)}^{ - \left( {2m + n} \right) - 1}}{{\left( { - 2} \right)}^n}}  \\
&\times \Bigg( \Bigg.
\frac{{{{\left( { - \frac{\varpi }{2}} \right)}^{2m + n + 1}}\Gamma \left( {k + 2m + 1} \right)}}{{\left( {2m + n + 1} \right){{\left( { - \frac{\varpi }{2} + z} \right)}^{k + 2m + 1}}}}\times{}_2{F_1}\left( {1,k + 2m + 1;2m + n + 2;\frac{{ - \varpi }}{{ - \varpi  + 2z}}} \right)\\
&-
\sum\limits_{i = 0}^{k - 1 - n} {\frac{\left( {k - 1 - n} \right)!}{i!{z^{k - n - i }}}}\Bigg( 
  {{\left( { - \frac{\varpi }{2}} \right)}^{2m + n + 1}} {{\left( { - \frac{\varpi }{2} + z} \right)}^{ - (2m + n + 1) - i}}  \\
  & \times  \Bigg(   
   {\gamma \bigg( {2m + n + 1 + i,\left( { - \frac{\varpi }{2} + z} \right)\ln \bigg( {\frac{{{\mu _{HD}}}}{{x{{\left( {\frac{z}{{z + 1}}} \right)}^k}\frac{\xi }{{\sqrt {\left( {1 + q\xi } \right)\left( {1 + \frac{\xi }{q}} \right)} }}}}} \bigg)} \bigg)}  \\
  &{ - \gamma \bigg( {2m + n + 1 , - \frac{\varpi }{2}\ln \bigg( {\frac{{{\mu _{HD}}}}{{x{{\left( {\frac{z}{{z + 1}}} \right)}^k}\frac{\xi }{{\sqrt {\left( {1 + q\xi } \right)\left( {1 + \frac{\xi }{q}} \right)} }}}}} \bigg)} \bigg){{\bigg( {\frac{{{\mu _{HD}}}}{{x{{\left( {\frac{z}{{z + 1}}} \right)}^k}\frac{\xi }{{\sqrt {\left( {1 + q\xi } \right)\left( {1 + \frac{\xi }{q}} \right)} }}}}} \bigg)}^{ - z}}} \\
  & \times {{{\bigg( {\ln \bigg( {\frac{{{\mu _{HD}}}}{{x{{\left( {\frac{z}{{z + 1}}} \right)}^k}\frac{\xi }{{\sqrt {\left( {1 + q\xi } \right)\left( {1 + \frac{\xi }{q}} \right)} }}}}} \bigg)} \bigg)}^i}}
  \Bigg) \Bigg) \Bigg).
\end{aligned}
\end{equation}

\section{Outage Probability based on IM/DD technique}
\label{AppenoutageprobabilityIMDD} 
From (\ref{pdfIMDD}), the outage probability for IM/DD technique case can be written as follows:
\begin{equation}
\begin{aligned}\label{outage probabilityIMDD}
 {F_\gamma }\left( x \right) &= \int_0^x {{f_\gamma }\left( \gamma  \right)d\gamma } \\
& = \frac{{{z^k}}}{{\Gamma \left( k \right)}}\xi {\bigg( {{{\left( {\frac{z}{{z + 1}}} \right)}^k}\frac{\xi }{{\sqrt {\left( {1 + q\xi } \right)\left( {1 + \frac{\xi }{q}} \right)} }}} \bigg)^z}{\left( {\frac{1}{\mu_{IM/DD} }} \right)^{\frac{z}{2}}}\sum\limits_{m = 0}^\infty  {\frac{1}{{{{\left( {m!} \right)}^2}}}{{\left( { - \frac{{\left( {1 - {q^2}} \right)\xi }}{{2q}}} \right)}^{2m}}    }\\   
&\times  { 
\sum\limits_{n = 0}^\infty  \binom{k-1}{n} }  \times {\left( { - \varpi } \right)^{ - \left( {2m + n} \right) - 1}}{\left( { - 2} \right)^n}\int_0^x {{\gamma ^{\frac{z}{2} - 1}}{{\bigg( {\ln \bigg( {\frac{{\sqrt {\frac{{{\mu _{IM/DD}}}}{\gamma }} }}{{{{\left( {\frac{z}{{z + 1}}} \right)}^k}\frac{\xi }{{\sqrt {\left( {1 + q\xi } \right)\left( {1 + \frac{\xi }{q}} \right)} }}}}} \bigg)} \bigg)}
^{k - 1 - n}}       }\\
&\times{
\gamma \bigg( {2m + n + 1, - \frac{\varpi }{2}\ln \bigg( {\frac{{\sqrt {\frac{{{\mu _{IM/DD}}}}{\gamma }} }}{{{{\left( {\frac{z}{{z + 1}}} \right)}^k}\frac{\xi }{{\sqrt {\left( {1 + q\xi } \right)\left( {1 + \frac{\xi }{q}} \right)} }}}}} \bigg)} \bigg)d\gamma } .
\end{aligned}
\end{equation}
Here, let $y\! =\! \ln\!\! \left( \!\!{\frac{{\sqrt {\frac{{{\mu _{IM/DD}}}}{\gamma }} }}{{{{\left( \!{\frac{z}{{z + 1}}}\! \right)}^k}\!\!\frac{\xi }{{\sqrt {\left( {1 + q\xi } \right)\left( {1 + \frac{\xi }{q}} \right)} }}}}}\!\! \right)$, then $\gamma \! =\!\!\!{\left(\!\!\! {\frac{{\sqrt {{\mu _{IM/DD}}} }}{{{{\left(\! {\frac{z}{{z + 1}}} \!\!\right)}^k}\!\!\frac{\xi }{\!{\sqrt {\left(\! {1 + q\xi } \!\right)\left(\! {1 + \frac{\xi }{q}}\! \right)} }}}}\!\exp\! \left( \!{ - y} \!\right)} \!\!\right)^{\!\!\!2}}$,
 {{$d\gamma  =  - 2
\exp \left( { - 2y} \right) \\
 \cdot{\left(\!\!\! {\frac{{\sqrt {{\mu _{IM/DD}}} }}{{{{\left(\! {\frac{z}{{z + 1}}} \!\!\right)}^k}\!\!\frac{\xi }{\!{\sqrt {\left(\! {1 + q\xi } \!\right)\left(\! {1 + \frac{\xi }{q}}\! \right)} }}}}} \!\!\right)^{\!\!\!2}}$}}dy, 
and the integral region of $y$ becomes: 
$\ln\!\! \left( \!\!{\frac{{\sqrt {\frac{{{\mu _{IM/DD}}}}{x }} }}{{{{\left( \!{\frac{z}{{z + 1}}}\! \right)}^k}\!\!\frac{\xi }{{\sqrt {\left( {1 + q\xi } \right)\left( {1 + \frac{\xi }{q}} \right)} }}}}}\!\! \right) < y < \infty $. 
Then, (\ref{outage probabilityIMDD}) can be expressed as follows:
\begin{equation} \label{outage probabilityIMDDIMDD}
\begin{aligned}
 {F_\gamma }\left( y \right) 
 &= \frac{{2{z^k}}}{{\Gamma \left( k \right)}}\xi \sum\limits_{m = 0}^\infty  {\frac{1}{{{{\left( {m!} \right)}^2}}}{{\left( { - \frac{{\left( {1 - {q^2}} \right)\xi }}{{2q}}} \right)}^{2m}}\sum\limits_{n = 0}^\infty  \binom{k-1}{n} } 
 {\left( { - \varpi } \right)^{ - \left( {2m + n} \right) - 1}}{\left( { - 2} \right)^n}\\
& \times
{\int_{\ln \left( {\frac{{\sqrt {\frac{{{\mu _{IM/DD}}}}{x}} }}{{{{\left( {\frac{z}{{z + 1}}} \right)}^k}\frac{\xi }{{\sqrt {\left( {1 + q\xi } \right)\left( {1 + \frac{\xi }{q}} \right)} }}}}} \right)}^\infty  {\gamma \left( {2m + n + 1, - \frac{\varpi }{2}y} \right)\exp \left( { - zy} \right){y^{k - 1 - n}}dy} }.
\end{aligned}
\end{equation}
The integral part in (\ref{outage probabilityIMDDIMDD}) can be divided into two parts (i.e., $0 < y < \infty $ and $0 < y < \ln \left( {\frac{{\sqrt {\frac{{{\mu _{IM/DD}}}}{x}} }}{{{{\left( {\frac{z}{{z + 1}}} \right)}^k}\frac{\xi }{{\sqrt {\left( {1 + q\xi } \right)\left( {1 + \frac{\xi }{q}} \right)} }}}}} \right)$).
Subsequently, using \cite[(Eq. (6.455.2)]{table}, the first integral part (i.e., $0 < y < \infty$) can be expressed as follows:
\begin{equation}
\begin{aligned}
\int_0^\infty  &{\gamma \left( {2m + n + 1, - \frac{\varpi }{2}y} \right)\exp \left( { - zy} \right){y^{k - 1 - n}}dy} \\
& = \frac{{{{\left( { - \frac{\varpi }{2}} \right)}^{2m + n + 1}}\Gamma \left( {k + 2m + 1} \right)}}{{\left( {2m + n + 1} \right){{\left( { - \frac{\varpi }{2} + z} \right)}^{k + 2m + 1}}}} \times {}_2{F_1}\left( {1,k + 2m + 1;2m + n + 2;\frac{{ - \varpi }}{{ - \varpi  + 2z}}} \right),
\end{aligned}
\end{equation}
and with the help of Theorem derived in Appendix \ref{Appendclosed}, the second integral part (i.e. $0 < y < \ln \left( {\frac{{\sqrt {\frac{{{\mu _{IM/DD}}}}{x}} }}{{{{\left( {\frac{z}{{z + 1}}} \right)}^k}\frac{\xi }{{\sqrt {\left( {1 + q\xi } \right)\left( {1 + \frac{\xi }{q}} \right)} }}}}} \right)$) can be expressed as follows:

\begin{equation}
\begin{aligned}
   &\int_0^{\ln \left( {\frac{{\sqrt {\frac{{{\mu _{IM/DD}}}}{x}} }}{{{{\left( {\frac{z}{{z + 1}}} \right)}^k}\frac{\xi }{{\sqrt {\left( {1 + q\xi } \right)\left( {1 + \frac{\xi }{q}} \right)} }}}}} \right)} {\gamma \left( {2m + n + 1, - \frac{\varpi }{2}y} \right)\exp \left( { - zy} \right){y^{k - 1 - n}}dy}  \\
  &= \sum\limits_{i = 0}^{k - 1 - n} {\frac{\left( {k - 1 - n} \right)!}{i!{z^{k - n - i }}}}\Bigg( \Bigg. 
  {{\left( { - \frac{\varpi }{2}} \right)}^{2m + n + 1}} {{\left( { - \frac{\varpi }{2} + z} \right)}^{ - (2m + n + 1) - i}}  \\
  & \times  \Bigg(   
   {\gamma \bigg( {2m + n + 1 + i,\left( { - \frac{\varpi }{2} + z} \right)\ln \bigg( {\frac{{\sqrt {\frac{{{\mu _{IM/DD}}}}{x}} }}{{{{\left( {\frac{z}{{z + 1}}} \right)}^k}\frac{\xi }{{\sqrt {\left( {1 + q\xi } \right)\left( {1 + \frac{\xi }{q}} \right)} }}}}}\bigg)} \bigg)}  \\
  &{ - \gamma \bigg( {2m + n + 1, - \frac{\varpi }{2}\ln \bigg( {\frac{{\sqrt {\frac{{{\mu _{IM/DD}}}}{x}} }}{{{{\left( {\frac{z}{{z + 1}}} \right)}^k}\frac{\xi }{{\sqrt {\left( {1 + q\xi } \right)\left( {1 + \frac{\xi }{q}} \right)} }}}}} \bigg)} \bigg){{\bigg( {\frac{{\sqrt {\frac{{{\mu _{IM/DD}}}}{x}} }}{{{{\left( {\frac{z}{{z + 1}}} \right)}^k}\frac{\xi }{{\sqrt {\left( {1 + q\xi } \right)\left( {1 + \frac{\xi }{q}} \right)} }}}}} \bigg)}^{ - z}}} \\
  & \times {{{\bigg( {\ln \bigg( {\frac{{\sqrt {\frac{{{\mu _{IM/DD}}}}{x}} }}{{{{\left( {\frac{z}{{z + 1}}} \right)}^k}\frac{\xi }{{\sqrt {\left( {1 + q\xi } \right)\left( {1 + \frac{\xi }{q}} \right)} }}}}}\bigg)} \bigg)}^i}}
  \Bigg) \Bigg).
\end{aligned}
\end{equation}

Consequently, the closed-form result of the outage probability of $\gamma$ for IM/DD technique case can be obtained as follows:
\begin{equation} \label{Appenoutage probabilityIMDDfinal}
\begin{aligned}
   {F_\gamma }\left( x  \right) &= \frac{{2{z^k}}}{{\Gamma \left( k \right)}}\xi {\sum\limits_{m = 0}^\infty  {\frac{1}{{{{\left( {m!} \right)}^2}}}\left( { - \frac{{\left( {1 - {q^2}} \right)\xi }}{{2q}}} \right)} ^{2m}}\sum\limits_{n = 0}^\infty \binom{k-1}{n} {{{\left( { - \varpi } \right)}^{ - \left( {2m + n} \right) - 1}}{{\left( { - 2} \right)}^n}}  \\
&\times \Bigg( 
\frac{{{{\left( { - \frac{\varpi }{2}} \right)}^{2m + n + 1}}\Gamma \left( {k + 2m + 1} \right)}}{{\left( {2m + n + 1} \right){{\left( { - \frac{\varpi }{2} + z} \right)}^{k + 2m + 1}}}}\times{}_2{F_1}\left( {1,k + 2m + 1;2m + n + 2;\frac{{ - \varpi }}{{ - \varpi  + 2z}}} \right)\\
&-
\sum\limits_{i = 0}^{k - 1 - n} {\frac{\left( {k - 1 - n} \right)!}{i!{z^{k - n - i }}}}\Bigg( 
  {{\left( { - \frac{\varpi }{2}} \right)}^{2m + n + 1}} {{\left( { - \frac{\varpi }{2} + z} \right)}^{ - (2m + n + 1) - i}}  \\
  & \times  \Bigg(   
   {\gamma \bigg( {2m + n + 1 + i,\left( { - \frac{\varpi }{2} + z} \right)\ln \bigg( {\frac{{\sqrt {\frac{{{\mu _{IM/DD}}}}{x}} }}{{{{\left( {\frac{z}{{z + 1}}} \right)}^k}\frac{\xi }{{\sqrt {\left( {1 + q\xi } \right)\left( {1 + \frac{\xi }{q}} \right)} }}}}} \bigg)} \bigg)}  \\
  &{ - \gamma \bigg( {2m + n + 1 , - \frac{\varpi }{2}\ln \bigg( {\frac{{\sqrt {\frac{{{\mu _{IM/DD}}}}{x}} }}{{{{\left( {\frac{z}{{z + 1}}} \right)}^k}\frac{\xi }{{\sqrt {\left( {1 + q\xi } \right)\left( {1 + \frac{\xi }{q}} \right)} }}}}} \bigg)} \bigg){{\bigg( {\frac{{\sqrt {\frac{{{\mu _{IM/DD}}}}{x}} }}{{{{\left( {\frac{z}{{z + 1}}} \right)}^k}\frac{\xi }{{\sqrt {\left( {1 + q\xi } \right)\left( {1 + \frac{\xi }{q}} \right)} }}}}} \bigg)}^{ - z}}} \\
  & \times {{{\bigg( {\ln \bigg( {\frac{{\sqrt {\frac{{{\mu _{IM/DD}}}}{x}} }}{{{{\left( {\frac{z}{{z + 1}}} \right)}^k}\frac{\xi }{{\sqrt {\left( {1 + q\xi } \right)\left( {1 + \frac{\xi }{q}} \right)} }}}}} \bigg)} \bigg)}^i}}
  \Bigg)\Bigg) \Bigg).
\end{aligned}
\end{equation}

\section{${\overline {BER} _{u,{\rm I}}}$ based on HD technique} 
\label{AppenBERHD}
For HD technique case, by substituting (\ref{pdfsnrHD}) and (\ref{outage probabilityHDfinal}) into (\ref{BERuI}), ${\overline {BER} _{u,{\rm I}}}$ can be expressed as follows:
\begin{equation}\label{BERuIHD}
\begin{aligned}
 {\overline {BER} _{u,{\rm I}}} &= \sum\limits_{l = 0}^\infty  {\frac{{{a_u}{{\left( { - \frac{{{b_u}}}{{{M_u}}}} \right)}^l}}}{{l!}}} {\textstyle{1 \over {{F_\gamma }\left( {{\textstyle{{{\mu _{HD}}} \over {{{\left( {\frac{z}{{z + 1}}} \right)}^k}\frac{\xi }{{\sqrt {\left( {1 + q\xi } \right)\left( {1 + \frac{\xi }{q}} \right)} }}}}}} \right) - {F_\gamma }\left( {{\gamma _T}} \right)}}}\frac{{2{z^k}}}{{\Gamma \left( k \right)}}\xi {\left( {{\textstyle{{{{\left( {\frac{z}{{z + 1}}} \right)}^k}\frac{\xi }{{\sqrt {\left( {1 + q\xi } \right)\left( {1 + \frac{\xi }{q}} \right)} }}} \over {{\mu _{HD}}}}}} \right)^z} \\
& \times
{\sum\limits_{m = 0}^\infty  {\frac{1}{{{{\left( {m!} \right)}^2}}}\left( { - \frac{{\left( {1 - {q^2}} \right)\xi }}{{2q}}} \right)} ^{2m}}\sum\limits_{n = 0}^\infty  {\binom{k-1}{n}{{\left( { - \varpi } \right)}^{ - \left( {2m + n} \right) - 1}}{{\left( { - 2} \right)}^n}}
 \\   
&  \times \int_{{\gamma _{{T_u}}}}^{{\textstyle{{{\mu _{HD}}} \over {{{\left( {\frac{z}{{z + 1}}} \right)}^k}\frac{\xi }{{\sqrt {\left( {1 + q\xi } \right)\left( {1 + \frac{\xi }{q}} \right)} }}}}}} {{\gamma ^{l + z - 1}}   {{{\bigg( {\ln \bigg( {{\textstyle{{{\mu _{HD}}} \over {{{\gamma\left( {\frac{z}{{z + 1}}} \right)}^k}\frac{\xi }{{\sqrt {\left( {1 + q\xi } \right)\left( {1 + \frac{\xi }{q}} \right)} }}}}}} \bigg)} \bigg)}^{k - 1 - n}}}    }\\
&\times{
\gamma \bigg( {2m + n + 1, - \frac{\varpi }{2}\ln \bigg( {{\textstyle{{{\mu _{HD}}} \over {{{\gamma\left( {\frac{z}{{z + 1}}} \right)}^k}\frac{\xi }{{\sqrt {\left( {1 + q\xi } \right)\left( {1 + \frac{\xi }{q}} \right)} }}}}}} \bigg)} \bigg)d\gamma } .
\end{aligned}
\end{equation}
Here, by letting {{$t = \ln \left( {\frac{{{\mu _{HD}}}}{{\gamma {{\left( {\frac{z}{{z + 1}}} \right)}^k}\frac{\xi }{{\sqrt {\left( {1 + q\xi } \right)\left( {1 + \frac{\xi }{q}} \right)} }}}}} \right)$, then $\gamma  = \frac{{{\mu _{HD}}}}{{{{\left( {\frac{z}{{z + 1}}} \right)}^k}\frac{\xi }{{\sqrt {\left( {1 + q\xi } \right)\left( {1 + \frac{\xi }{q}} \right)} }}}}\exp \left( { - t} \right)$ and $d\gamma={\textstyle{{ - {\mu _{HD}}} \over {{{\left( {\frac{z}{{z + 1}}} \right)}^k}\frac{\xi }{{\sqrt {\left( {1 + q\xi } \right)\left( {1 + \frac{\xi }{q}} \right)} }}}}}
\cdot \exp \left( { - t} \right)dt$}}, and the integral region of $t$ becomes $0 < t < \ln \left( {\frac{{{\mu _{HD}}}}{{{\gamma _{{T_u}}}{{\left( {\frac{z}{{z + 1}}} \right)}^k}\frac{\xi }{{\sqrt {\left( {1 + q\xi } \right)\left( {1 + \frac{\xi }{q}} \right)} }}}}} \right)$.
Thus, (\ref{BERuIHD}) can be rewritten as follows:
\begin{equation}\label{BERIfinal}
\begin{aligned}
 {\overline {BER} _{u,{\rm I}}} &= \sum\limits_{l = 0}^\infty  {\frac{{{a_u}{{\left( { - \frac{{{b_u}}}{{{M_u}}}} \right)}^l}}}{{l!}}} {\textstyle{1 \over {{F_\gamma }\left( {{\textstyle{{{\mu _{HD}}} \over {{{\left( {\frac{z}{{z + 1}}} \right)}^k}\frac{\xi }{{\sqrt {\left( {1 + q\xi } \right)\left( {1 + \frac{\xi }{q}} \right)} }}}}}} \right) - {F_\gamma }\left( {{\gamma _T}} \right)}}}\frac{{2{z^k}}}{{\Gamma \left( k \right)}}\xi {\left( {{\textstyle{{{\mu _{HD}}} \over {{{\left( {\frac{z}{{z + 1}}} \right)}^k}\frac{\xi }{{\sqrt {\left( {1 + q\xi } \right)\left( {1 + \frac{\xi }{q}} \right)} }}}}}} \right)^l} \\
& \times
{\sum\limits_{m = 0}^\infty  {\frac{1}{{{{\left( {m!} \right)}^2}}}\left( { - \frac{{\left( {1 - {q^2}} \right)\xi }}{{2q}}} \right)} ^{2m}}\sum\limits_{n = 0}^\infty  {\binom{k-1}{n}{{\left( { - \varpi } \right)}^{ - \left( {2m + n} \right) - 1}}{{\left( { - 2} \right)}^n}}
 \\   
&  \times \int_{{0}}^{{\ln \left( {{\textstyle{{{\mu _{HD}}} \over {{\gamma _{{T_u}}}{{\left( {\frac{z}{{z + 1}}} \right)}^k}\frac{\xi }{{\sqrt {\left( {1 + q\xi } \right)\left( {1 + \frac{\xi }{q}} \right)} }}}}}} \right)}} {\gamma \left( {2m + n + 1, - \frac{\varpi }{2}t} \right)\exp \left( { - t\left( {l + z} \right)} \right){t^{k - 1 - n}}dt} .
\end{aligned}
\end{equation}
With (\ref{BERIfinal}), by applying the theorem  derived in Appendix \ref{Appendclosed}, the integral part in (\ref{BERIfinal}) can be expressed as follows:
\begin{equation}
\begin{aligned}
   &\int_0^{\ln \left( {{\textstyle{{{\mu _{HD}}} \over {{\gamma _{{T_u}}}{{\left( {\frac{z}{{z + 1}}} \right)}^k}\frac{\xi }{{\sqrt {\left( {1 + q\xi } \right)\left( {1 + \frac{\xi }{q}} \right)} }}}}}} \right)} {\gamma \left( {2m + n + 1, - \frac{\varpi }{2}t} \right)} \exp \left( { - t\left( {l + z} \right)} \right){t^{k - 1 - n}}dt \\
  &= \sum\limits_{i = 0}^{k - 1 - n}{\frac{{\left( {k - 1 - n} \right)!}}{{i!{{\left( {l + z} \right)}^{k - n - i}}}}} 
\Bigg(  
  {{\left( { - \frac{\varpi }{2}} \right)}^{-2m+n+1}} {{\left( { - \frac{\varpi }{2}  +l+z} \right)}^{ - (2m + n + 1) - i}}  \\
  & \times    
   {\gamma \bigg( {2m + n + 1 + i,\left( { - \frac{\varpi }{2} + l+ z} \right)\ln \bigg( {{{{{\mu _{HD}}} \over {{\gamma _{{T_u}}}{{\left( {\frac{z}{{z + 1}}} \right)}^k}\frac{\xi }{{\sqrt {\left( {1 + q\xi } \right)\left( {1 + \frac{\xi }{q}} \right)} }}}}}} \bigg)} \bigg)}  \\
  &{ - \gamma \bigg( {2m + n + 1 , - \frac{\varpi }{2}\ln \bigg( {{{{{\mu _{HD}}} \over {{\gamma _{{T_u}}}{{\left( {\frac{z}{{z + 1}}} \right)}^k}\frac{\xi }{{\sqrt {\left( {1 + q\xi } \right)\left( {1 + \frac{\xi }{q}} \right)} }}}}}} \bigg)} \bigg){{\bigg( {{{{{\mu _{HD}}} \over {{\gamma _{{T_u}}}{{\left( {\frac{z}{{z + 1}}} \right)}^k}\frac{\xi }{{\sqrt {\left( {1 + q\xi } \right)\left( {1 + \frac{\xi }{q}} \right)} }}}}}} \bigg)}^{ - l-z}}} \\
  & \times {{{\bigg( {\ln \bigg( {\frac{{{\mu _{HD}}}}{{{\gamma _{{T_u}}}{{\left( {\frac{z}{{z + 1}}} \right)}^k}\frac{\xi }{{\sqrt {\left( {1 + q\xi } \right)\left( {1 + \frac{\xi }{q}} \right)} }}}}} \bigg)} \bigg)}^i}}
  \Bigg).
\end{aligned}
\end{equation}

Consequently, the closed-form result of ${\overline {BER} _{u,{\rm I}}}$ for HD technique case can be obtained as follows:
\begin{equation}  \label{BERuIHDclosed}
\begin{aligned}
 {\overline {BER} _{u,{\rm I}}} &= \sum\limits_{l = 0}^\infty  {\frac{{{a_u}{{\left( { - \frac{{{b_u}}}{{{M_u}}}} \right)}^l}}}{{l!}} \frac{1}{{{F_\gamma }\left( {{{\textstyle{{{\mu _{HD}}} \over {{{\left( {\frac{z}{{z + 1}}} \right)}^k}\frac{\xi }{{\sqrt {\left( {1 + q\xi } \right)\left( {1 + \frac{\xi }{q}} \right)} }}}}}}} \right) - {F_\gamma }\left( {{\gamma _T}} \right)}}} \frac{{2{z^k}}}{{\Gamma \left( k \right)}}\xi {\left( {\frac{{{\mu _{HD}}}}{{{{\left( {\frac{z}{{z + 1}}} \right)}^k}\frac{\xi }{{\sqrt {\left( {1 + q\xi } \right)\left( {1 + \frac{\xi }{q}} \right)} }}}}} \right)^l}\\
  &\times 
{\sum\limits_{m = 0}^\infty  {\frac{1}{{{{\left( {m!} \right)}^2}}}\left( { - \frac{{\left( {1 - {q^2}} \right)\xi }}{{2q}}} \right)} ^{2m}}\sum\limits_{n = 0}^\infty  \binom{k-1}{n} {\left( { - \varpi } \right)^{ - \left( {2m + n} \right) - 1}}{\left( { - 2} \right)^n}\\
  &\times
  \sum\limits_{i = 0}^{k - 1 - n}{\frac{{\left( {k - 1 - n} \right)!}}{{i!{{\left( {l + z} \right)}^{k - n - i}}}}} 
\Bigg(  
  {{\left( { - \frac{\varpi }{2}} \right)}^{-2m+n+1}} {{\left( { - \frac{\varpi }{2}  +l+z} \right)}^{ - (2m + n + 1) - i}}  \\
  & \times    
   {\gamma \bigg( {2m + n + 1 + i,\left( { - \frac{\varpi }{2} + l+ z} \right)\ln \bigg( {{{{{\mu _{HD}}} \over {{\gamma _{{T_u}}}{{\left( {\frac{z}{{z + 1}}} \right)}^k}\frac{\xi }{{\sqrt {\left( {1 + q\xi } \right)\left( {1 + \frac{\xi }{q}} \right)} }}}}}} \bigg)} \bigg)}  \\
  &{ - \gamma \bigg( {2m + n + 1 , - \frac{\varpi }{2}\ln \bigg( {{{{{\mu _{HD}}} \over {{\gamma _{{T_u}}}{{\left( {\frac{z}{{z + 1}}} \right)}^k}\frac{\xi }{{\sqrt {\left( {1 + q\xi } \right)\left( {1 + \frac{\xi }{q}} \right)} }}}}}} \bigg)} \bigg){{\bigg( {{{{{\mu _{HD}}} \over {{\gamma _{{T_u}}}{{\left( {\frac{z}{{z + 1}}} \right)}^k}\frac{\xi }{{\sqrt {\left( {1 + q\xi } \right)\left( {1 + \frac{\xi }{q}} \right)} }}}}}} \bigg)}^{ - l-z}}} \\
  & \times {{{\bigg( {\ln \bigg( {\frac{{{\mu _{HD}}}}{{{\gamma _{{T_u}}}{{\left( {\frac{z}{{z + 1}}} \right)}^k}\frac{\xi }{{\sqrt {\left( {1 + q\xi } \right)\left( {1 + \frac{\xi }{q}} \right)} }}}}} \bigg)} \bigg)}^i}}
  \Bigg).
\end{aligned}
\end{equation}

\section{${\overline {BER} _{u,{\rm I}}}$ based on IM/DD technique}
\label{AppenBERIMDD}
For IM/DD technique case, by substituting (\ref{pdfIMDD}) and (\ref{outage probabilityIMDDfinal}) into (\ref{BERuI}), ${\overline {BER} _{u,{\rm I}}}$ can be expressed as follows:
\begin{equation}\label{BERuIIMDD}
\begin{aligned}
 {\overline {BER} _{u,{\rm I}}} &= \sum\limits_{l = 0}^\infty  {\frac{{{a_u}{{\left( { - \frac{{{b_u}}}{{{M_u}}}} \right)}^l}}}{{l!}}} {\textstyle{1 \over {{F_\gamma }\left( {\frac{{{\mu _{IM/DD}}}}{{{{\left( {{{\left( {\frac{z}{{z + 1}}} \right)}^k}\frac{\xi }{{\sqrt {\left( {1 + q\xi } \right)\left( {1 + \frac{\xi }{q}} \right)} }}} \right)}^2}}}} \right) - {F_\gamma }\left( {{\gamma _T}} \right)}}}\frac{{{z^k}}}{{\Gamma \left( k \right)}}\xi {\left( {\frac{{{{\left( {\frac{z}{{z + 1}}} \right)}^k}\frac{\xi }{{\sqrt {\left( {1 + q\xi } \right)\left( {1 + \frac{\xi }{q}} \right)} }}}}{{\sqrt {{\mu _{IM/DD}}} }}} \right)^z} \\
& \times
{\sum\limits_{m = 0}^\infty  {\frac{1}{{{{\left( {m!} \right)}^2}}}\left( { - \frac{{\left( {1 - {q^2}} \right)\xi }}{{2q}}} \right)} ^{2m}}\sum\limits_{n = 0}^\infty  {\binom{k-1}{n}{{\left( { - \varpi } \right)}^{ - \left( {2m + n} \right) - 1}}{{\left( { - 2} \right)}^n}}
 \\   
&  \times \int_{{\gamma _{{T_u}}}}^{\frac{{{\mu _{IM/DD}}}}{{{{\left( {{{\left( {\frac{z}{{z + 1}}} \right)}^k}\frac{\xi }{{\sqrt {\left( {1 + q\xi } \right)\left( {1 + \frac{\xi }{q}} \right)} }}} \right)}^2}}}}
 {{\gamma ^{l + {\frac{z}{2}} - 1}}   {{{\bigg( {\ln \bigg( {\frac{{\sqrt {\frac{{{\mu _{IM/DD}}}}{\gamma }} }}{{{{\left( {\frac{z}{{z + 1}}} \right)}^k}\frac{\xi }{{\sqrt {\left( {1 + q\xi } \right)\left( {1 + \frac{\xi }{q}} \right)} }}}}} \bigg)} \bigg)}^{k - 1 - n}}}    }\\
&\times{
\gamma \bigg( {2m + n + 1, - \frac{\varpi }{2}\ln \bigg( {\frac{{\sqrt {\frac{{{\mu _{IM/DD}}}}{\gamma }} }}{{{{\left( {\frac{z}{{z + 1}}} \right)}^k}\frac{\xi }{{\sqrt {\left( {1 + q\xi } \right)\left( {1 + \frac{\xi }{q}} \right)} }}}}} \bigg)} \bigg)d\gamma } .
\end{aligned}
\end{equation}
By letting $t \!\!=\!\!\ln\!\! \left(\!\!\! {\frac{{\sqrt {\frac{{{\mu _{IM/DD}}}}{\gamma }} }}{{{{\left( \!{\frac{z}{{z + 1}}} \!\right)}^k}\!\!\!\frac{\xi }{\!\!\!{\sqrt {\left( {1 + q\xi } \right)\left( {1 + \frac{\xi }{q}} \right)} }}}}} \!\!\right)$, then
 $\gamma \! \!= \!\!{\left(\!\!\! {\frac{{\sqrt {{\mu _{IM/DD}}} }}{{{{\left( \!\!{\frac{z}{{z + 1}}}\!\! \right)}^k}\!\!\!\frac{\xi }{{\!\!\!\sqrt {\left(\! {1 + q\xi }\! \right)\left( \!\!{1 + \frac{\xi }{q}} \!\!\right)} }}}}\!\exp \!\left( \!{ - t} \!\right)} \!\!\! \right)^{\!\!\!\! 2}}$ 
and $d\gamma  \!\!=\!\! {\left(\!\!\!  {\frac{{\sqrt {{{{\mu _{IM/DD}}}}} }}{{{{\left( \!{\frac{z}{{z + 1}}} \!\right)}^k}\!\!\!\frac{\xi }{\!\!\!{\sqrt {\left( {1 + q\xi } \right)\left( {1 + \frac{\xi }{q}} \right)} }}}}} \!\!\! \right)^{\!\!\! 2}} \\
\cdot \left( { - 2\exp \left( { - 2t} \right)} \right)dt$, and the integral region of $t$ becomes $0 < t < \ln \left( {\frac{{\sqrt {\frac{{{\mu _{IM/DD}}}}{\gamma_{Tu} }} }}{{{{\left( {\frac{z}{{z + 1}}} \right)}^k}\frac{\xi }{{\sqrt {\left( {1 + q\xi } \right)\left( {1 + \frac{\xi }{q}} \right)} }}}}} \right)$.
Thus, (\ref{BERuIIMDD}) is expressed as follows:
\begin{equation}\label{BERuIIMDD2}
\begin{aligned}
 {\overline {BER} _{u,{\rm I}}} &= \sum\limits_{l = 0}^\infty  {\frac{{{a_u}{{\left( { - \frac{{{b_u}}}{{{M_u}}}} \right)}^l}}}{{l!}}}{{\textstyle{1 \over {{F_\gamma }\left( {\frac{{{\mu _{IM/DD}}}}{{{{\left( {{{\left( {\frac{z}{{z + 1}}} \right)}^k}\frac{\xi }{{\sqrt {\left( {1 + q\xi } \right)\left( {1 + \frac{\xi }{q}} \right)} }}} \right)}^2}}}} \right) - {F_\gamma }\left( {{\gamma _T}} \right)}}}} \frac{{2{z^k}}}{{\Gamma \left( k \right)}}\xi {\left( {\frac{{\sqrt {{\mu _{IM/DD}}} }}{{{{\left( {\frac{z}{{z + 1}}} \right)}^k}\frac{\xi }{{\sqrt {\left( {1 + q\xi } \right)\left( {1 + \frac{\xi }{q}} \right)} }}}}} \right)^{2l}} \\
& \times
{\sum\limits_{m = 0}^\infty  {\frac{1}{{{{\left( {m!} \right)}^2}}}\left( { - \frac{{\left( {1 - {q^2}} \right)\xi }}{{2q}}} \right)} ^{2m}}\sum\limits_{n = 0}^\infty  {\binom{k-1}{n}{{\left( { - \varpi } \right)}^{ - \left( {2m + n} \right) - 1}}{{\left( { - 2} \right)}^n}}
 \\   
&  \times \int_{{0}}^{{\ln \left( {\frac{{\sqrt {\frac{{{\mu _{IM/DD}}}}{{{\gamma _{{T_u}}}}}} }}{{{{\left( {\frac{z}{{z + 1}}} \right)}^k}\frac{\xi }{{\sqrt {\left( {1 + q\xi } \right)\left( {1 + \frac{\xi }{q}} \right)} }}}}} \right)}} {\gamma \left( {2m + n + 1, - \frac{\varpi }{2}t} \right)\exp \left( { - t\left( {2l + z} \right)} \right){t^{k - 1 - n}}dt} .
\end{aligned}
\end{equation}
Then, by applying the theorem derived in Appendix \ref{Appendclosed}, the integral part in (\ref{BERuIIMDD2}) can be obtained as follows: 
\begin{equation}
\begin{aligned}
   & \int_{{0}}^{{\ln \left( {\frac{{\sqrt {\frac{{{\mu _{IM/DD}}}}{{{\gamma _{{T_u}}}}}} }}{{{{\left( {\frac{z}{{z + 1}}} \right)}^k}\frac{\xi }{{\sqrt {\left( {1 + q\xi } \right)\left( {1 + \frac{\xi }{q}} \right)} }}}}} \right)}} {\gamma \left( {2m + n + 1, - \frac{\varpi }{2}t} \right)\exp \left( { - t\left( {2l + z} \right)} \right){t^{k - 1 - n}}dt} \\
  &= \sum\limits_{i = 0}^{k - 1 - n}{\frac{{\left( {k - 1 - n} \right)!}}{{i!{{\left( {2l + z} \right)}^{k - n - i}}}}} 
\Bigg( \Bigg. 
  {{\left( { - \frac{\varpi }{2}} \right)}^{2m + n + 1}} {{\left( { - \frac{\varpi }{2} +2l+ z} \right)}^{ - (2m + n + 1) - i}}  \\
  & \times    
   {\gamma \bigg( {2m + n + 1 + i,\left( { - \frac{\varpi }{2} +2l+ z} \right)\ln \bigg( {\frac{{\sqrt {\frac{{{\mu _{IM/DD}}}}{{{\gamma _{{T_u}}}}}} }}{{{{\left( {\frac{z}{{z + 1}}} \right)}^k}\frac{\xi }{{\sqrt {\left( {1 + q\xi } \right)\left( {1 + \frac{\xi }{q}} \right)} }}}}} \bigg)} \bigg)}  \\
  &{ - \gamma \bigg( {2m + n + 1 , - \frac{\varpi }{2}\ln \bigg( {\frac{{\sqrt {\frac{{{\mu _{IM/DD}}}}{{{\gamma _{{T_u}}}}}} }}{{{{\left( {\frac{z}{{z + 1}}} \right)}^k}\frac{\xi }{{\sqrt {\left( {1 + q\xi } \right)\left( {1 + \frac{\xi }{q}} \right)} }}}}} \bigg)} \bigg){{\bigg( {\frac{{\sqrt {\frac{{{\mu _{IM/DD}}}}{{{\gamma _{{T_u}}}}}} }}{{{{\left( {\frac{z}{{z + 1}}} \right)}^k}\frac{\xi }{{\sqrt {\left( {1 + q\xi } \right)\left( {1 + \frac{\xi }{q}} \right)} }}}}} \bigg)}^{{ - \left( {2l + z} \right)}}}} \\
  & \times {{{\bigg( {\ln \bigg({\frac{{\sqrt {\frac{{{\mu _{IM/DD}}}}{{{\gamma _{{T_u}}}}}} }}{{{{\left( {\frac{z}{{z + 1}}} \right)}^k}\frac{\xi }{{\sqrt {\left( {1 + q\xi } \right)\left( {1 + \frac{\xi }{q}} \right)} }}}}} \bigg)} \bigg)}^i}}
 \Bigg. \Bigg).
\end{aligned}
\end{equation}

Thus, the closed-form result of ${\overline {BER} _{u,{\rm I}}}$ for IM/DD technique case can be obtained as follows: 
\begin{equation}  \label{BERuIIMDDclosed}
\begin{aligned}
{\overline {BER} _{u,{\rm I}}} &= \sum\limits_{l = 0}^\infty  {\frac{{{a_u}{{\left( { - \frac{{{b_u}}}{{{M_u}}}} \right)}^l}}}{{l!}}}{{\textstyle{1 \over {{F_\gamma }\left( {\frac{{{\mu _{IM/DD}}}}{{{{\left( {{{\left( {\frac{z}{{z + 1}}} \right)}^k}\frac{\xi }{{\sqrt {\left( {1 + q\xi } \right)\left( {1 + \frac{\xi }{q}} \right)} }}} \right)}^2}}}} \right) - {F_\gamma }\left( {{\gamma _T}} \right)}}}} \frac{{2{z^k}}}{{\Gamma \left( k \right)}}\xi {\left( {\frac{{\sqrt {{\mu _{IM/DD}}} }}{{{{\left( {\frac{z}{{z + 1}}} \right)}^k}\frac{\xi }{{\sqrt {\left( {1 + q\xi } \right)\left( {1 + \frac{\xi }{q}} \right)} }}}}} \right)^{2l}} \\
  &\times 
{\sum\limits_{m = 0}^\infty  {\frac{1}{{{{\left( {m!} \right)}^2}}}\left( { - \frac{{\left( {1 - {q^2}} \right)\xi }}{{2q}}} \right)} ^{2m}}\sum\limits_{n = 0}^\infty  \binom{k-1}{n} {\left( { - \varpi } \right)^{ - \left( {2m + n} \right) - 1}}{\left( { - 2} \right)^n}\\
  &\times
\sum\limits_{i = 0}^{k - 1 - n}{\frac{{\left( {k - 1 - n} \right)!}}{{i!{{\left( {2l + z} \right)}^{k - n - i}}}}} 
\Bigg(  
  {{\left( { - \frac{\varpi }{2}} \right)}^{2m + n + 1}} {{\left( { - \frac{\varpi }{2} +2l+ z} \right)}^{ - (2m + n + 1) - i}}  \\
  & \times    
   {\gamma \bigg( {2m + n + 1 + i,\left( { - \frac{\varpi }{2} +2l+ z} \right)\ln \bigg( {\frac{{\sqrt {\frac{{{\mu _{IM/DD}}}}{{{\gamma _{{T_u}}}}}} }}{{{{\left( {\frac{z}{{z + 1}}} \right)}^k}\frac{\xi }{{\sqrt {\left( {1 + q\xi } \right)\left( {1 + \frac{\xi }{q}} \right)} }}}}} \bigg)} \bigg)}  \\
  &{ - \gamma \bigg( {2m + n + 1 , - \frac{\varpi }{2}\ln \bigg( {\frac{{\sqrt {\frac{{{\mu _{IM/DD}}}}{{{\gamma _{{T_u}}}}}} }}{{{{\left( {\frac{z}{{z + 1}}} \right)}^k}\frac{\xi }{{\sqrt {\left( {1 + q\xi } \right)\left( {1 + \frac{\xi }{q}} \right)} }}}}} \bigg)} \bigg){{\bigg( {\frac{{\sqrt {\frac{{{\mu _{IM/DD}}}}{{{\gamma _{{T_u}}}}}} }}{{{{\left( {\frac{z}{{z + 1}}} \right)}^k}\frac{\xi }{{\sqrt {\left( {1 + q\xi } \right)\left( {1 + \frac{\xi }{q}} \right)} }}}}} \bigg)}^{{ - \left( {2l + z} \right)}}}} \\
  & \times {{{\bigg( {\ln \bigg({\frac{{\sqrt {\frac{{{\mu _{IM/DD}}}}{{{\gamma _{{T_u}}}}}} }}{{{{\left( {\frac{z}{{z + 1}}} \right)}^k}\frac{\xi }{{\sqrt {\left( {1 + q\xi } \right)\left( {1 + \frac{\xi }{q}} \right)} }}}}} \bigg)} \bigg)}^i}}
  \Bigg).
\end{aligned}
\end{equation}

{
\section{Theorem}
\label{Appendclosed}
\textit{Proof:} 
\begin{equation}  
\begin{aligned}
&\int_0^d {\gamma \left( {a,bx} \right)\exp \left( { - cx} \right){x^k}dx} \\
& = \sum\limits_{i = o}^k {\frac{{k!}}{{i!{c^{k - i + 1}}}}} \left( {{b^a}{{\left( {b + c} \right)}^{ - a - i}}} \right.\gamma \left( {a + i,\left( {b + c} \right)d} \right) \left. {- \gamma \left( {a,bd} \right)\exp \left( { - cd} \right){d^i}} \right) .
\end{aligned}
\end{equation}
By using the definition of integration by parts \cite[Eq. (2.02.5)]{table}, $\int_0^d {\gamma \left( {a,bx} \right)\exp \left( { - cx} \right){x^k}dx}$ can be rewritten as follows:
\begin{equation} \label{th2}
\begin{aligned}
&\int_0^d {\frac{\partial }{{\partial x}}\left( {\gamma \left( {a,bx} \right)\exp \left( { - cx} \right){x^k}} \right)dx}  \\
&= \int_0^d {\frac{\partial }{{\partial x}}\left( {\gamma \left( {a,bx} \right)\exp \left( { - cx} \right)} \right){x^k}dx} + \int_0^d {\gamma \left( {a,bx} \right)\exp \left( { - cx} \right)\frac{\partial }{{\partial x}}\left( {{x^k}} \right)dx} .
\end{aligned}
\end{equation}
In (\ref{th2}), by applying the product rule and \cite[Eq. (8.356.4)]{table}, we can express (\ref{th2}) as follows: 
\begin{equation} \label{th3} 
\begin{aligned}
&\int_0^d {\frac{\partial }{{\partial x}}\left( {\gamma \left( {a,bx} \right)\exp \left( { - cx} \right){x^k}} \right)dx} \\
& = {b^a}\int_0^d {\exp \left( { - \left( {b + c} \right)x} \right){x^{a + k- 1 }}dx} \\
& + \int_0^d {\gamma \left( {a,bx} \right)\left( { - c\exp \left( { - cx} \right)} \right){x^k}dx}  + k\int_0^d {\gamma \left( {a,bx} \right)\exp \left( { - cx} \right){x^{k-1}}dx}  .
\end{aligned}
\end{equation}
Then, by arranging (\ref{th3}) in the desired form, we rewrite (\ref{th3}) as follows:
\begin{equation} \label{th4} 
\begin{aligned}
&\int_0^d {\gamma \left( {a,bx} \right)\exp \left( { - cx} \right){x^k}dx} \\
& = \frac{{{b^a}}}{c}\int_0^d {\exp \left( { - \left( {b + c} \right)x} \right){x^{a  + k- 1}}dx} \\
&  - \frac{1}{c}\int_0^d {\frac{\partial }{{\partial x}}\left( {\gamma \left( {a,bx} \right)\exp \left( { - cx} \right){x^k}} \right)dx}  + \frac{k}{c}\int_0^d {\gamma \left( {a,bx} \right)\exp \left( { - cx} \right){x^{k - 1}}dx}.
\end{aligned}
\end{equation}
The final term in (\ref{th4}) can be expressed using (\ref{th4}) as follows: 
\begin{equation} \label{th6} 
\begin{aligned}
&\int_0^d {\gamma \left( {a,bx} \right)\exp \left( { - cx} \right){x^{k - 1}}dx} \\
& = \frac{{{b^a}}}{c}\int_0^d {\exp \left( { - \left( {b + c} \right)x} \right){x^{a + k - 2}}dx} \\
& - \frac{1}{c}\int_0^d {\frac{\partial }{{\partial x}}\left( {\gamma \left( {a,bx} \right)\exp \left( { - cx} \right){x^{k - 1}}} \right)dx}  + \frac{{k - 1}}{c}\int_0^d {\gamma \left( {a,bx} \right)\exp \left( { - cx} \right){x^{k - 2}}dx} .
\end{aligned}
\end{equation}
Consequently, by repeating the above process until the power of $x$ becomes $0$ (i.e., this expression is valid when $k$ is an integer), we obtain the expression as follows: 
\begin{equation} \label{th7} 
\begin{aligned}
&\int_0^d {\gamma \left( {a,bx} \right)\exp \left( { - cx} \right){x^k}dx}  \\
&= \frac{1}{c}\left[ {{b^a}\int_0^d {\exp \left( { - \left( {b + c} \right)x} \right){x^{a + k - 1}}dx}  - \int_0^d {\frac{\partial }{{\partial x}}\left( {\gamma \left( {a,bx} \right)\exp \left( { - cx} \right){x^k}} \right)dx} } \right] \\
&+ \frac{k}{{{c^2}}}\left[ {{b^a}\int_0^d {\exp \left( { - \left( {b + c} \right)x} \right){x^{a + k - 2}}dx}  - \int_0^d {\frac{\partial }{{\partial x}}\left( {\gamma \left( {a,bx} \right)\exp \left( { - cx} \right){x^{k - 1}}} \right)dx} } \right]\\
&+ \frac{k\left(k-1 \right)}{{{c^3}}}\left[ {{b^a}\int_0^d {\exp \left( { - \left( {b + c} \right)x} \right){x^{a + k - 3}}dx}  - \int_0^d {\frac{\partial }{{\partial x}}\left( {\gamma \left( {a,bx} \right)\exp \left( { - cx} \right){x^{k - 1}}} \right)dx} } \right]\\
&\cdots + \frac{{k\left( {k - 1} \right) \cdots 1}}{{{c^{k + 1}}}}\left[ {{b^a}\int_0^d {\exp \left( { - \left( {b + c} \right)x} \right){x^{a -1}}dx}  - \int_0^d {\frac{\partial }{{\partial x}}\left( {\gamma \left( {a,bx} \right)\exp \left( { - cx} \right)} \right)dx} } \right] .
\end{aligned}
\end{equation}

In (\ref{th7}), we apply \cite[Eq. (3.351.1)]{table}, and then, with some algebraic simplifications, we obtain the final closed-form expression as follows:
\begin{equation} \label{th5} 
\begin{aligned}
&\int_0^d {\gamma \left( {a,bx} \right)\exp \left( { - cx} \right){x^k}dx}  \\
& = \sum\limits_{i = o}^k {\frac{{k!}}{{i!{c^{k - i + 1}}}}} \left( {{b^a}{{\left( {b + c} \right)}^{ - a - i}}} \right.\gamma \left( {a + i,\left( {b + c} \right)d} \right) \left. {- \gamma \left( {a,bd} \right)\exp \left( { - cd} \right){d^i}} \right)
\end{aligned}
\end{equation}
}
} 
Therefore, for $\alpha  < \beta $, we can generalize the result above as follows:
\begin{equation}
\begin{aligned}
&\int_\alpha ^\beta  {\gamma \left( {a,bx} \right)\exp \left( { - cx} \right){x^k}dx} \\
&  = \sum\limits_{i = o}^k {\frac{{k!}}{{i!{c^{k - i+ 1}}}}} \left[ {{b^a}} \right.{\left( {b + c} \right)^{ - a - i}}\left( {\gamma \left( {a + i,\left( {b + c} \right)\beta } \right) - \gamma \left( {a + i,\left( {b + c} \right)\alpha } \right)} \right) \\
&\qquad  - \left. {\gamma \left( {a,b\beta } \right)\exp \left( { - c\beta } \right){\beta ^i} + \gamma \left( {a,b\alpha } \right)\exp \left( { - c\alpha } \right)\alpha^i } \right].
\end{aligned}
\end{equation}

\bibliographystyle{ieeetran}
\bibliography{IEEEabrv,mybib}

\end{document}